\documentclass[bj]{imsart}

\RequirePackage{amsthm,amsmath,amsfonts,amssymb}
\RequirePackage[numbers]{natbib}

\RequirePackage[colorlinks,citecolor=blue,urlcolor=blue]{hyperref}
\RequirePackage{graphicx}

\usepackage{algorithm}
\usepackage{algorithmicx}
\usepackage[noend]{algpseudocode}
\usepackage{todonotes}
\usepackage{tikz}

\startlocaldefs
\theoremstyle{plain}
\newtheorem{theorem}{Theorem}
\newtheorem{proposition}{Proposition}
\newtheorem{lemma}{Lemma}
\newtheorem{corollary}{Corollary}

\theoremstyle{remark}
\newtheorem{assumption}{Assumption}
\newtheorem{remark}{Remark}

\makeatletter
\def\namedlabel#1#2{\begingroup
    #2%
    \def\@currentlabel{#2}%
    \phantomsection\label{#1}\endgroup
}
\makeatother

\DeclareMathOperator*{\mydiag}{\mathrm{diag}}
\DeclareMathOperator*{\argmin}{\mathrm{arg\,min}}

\newcommand{\boundborel}{\mathcal{B}}
\newcommand{\Lmax}{L}

\newcommand{\E}{\mathbb{E}}
\newcommand{\F}{\mathcal{F}}

\newcommand{\G}{\mathcal{G}}
\newcommand{\I}{\mathbb{I}}
\newcommand{\calI}{\mathcal{I}}

\newcommand{\N}{\mathbb{N}}
\newcommand{\calN}{\mathcal{N}}
\newcommand{\Ntot}{S}

\renewcommand{\P}{\mathbb{P}}

\newcommand{\R}{\mathbb{R}}
\newcommand{\rhomax}{\rho^{\mathrm{max}}}
\newcommand{\sign}{\mathrm{sgn}}

\newcommand{\ud}{\mathrm{d}}

\newcommand{\X}{\mathbb{X}}
\newcommand{\Y}{\mathbb{Y}}

\newcommand{\asto}[1]{\xrightarrow[#1]{\mathrm{a.s.}}}
\newcommand{\indist}[1]{\xrightarrow[#1]{\mathrm{D}}}
\newcommand{\inprob}[1]{\xrightarrow[#1]{\mathbb{P}}}

\newcommand{\vt}{\boldsymbol{t}}

\newcommand{\ovarphi}{\overline{\varphi}}
\newcommand{\defeq}{:=}

\newcommand{\vvarphi}{\boldsymbol{\varphi}}
\newcommand{\bGamma}{\boldsymbol{\Gamma}}

\definecolor{myhighlight}{rgb}{0.89, 0.0, 0.13}
\endlocaldefs

\begin{document}

\begin{frontmatter}
\title{Multilevel Bootstrap Particle Filter}
\runtitle{Multilevel Bootstrap Particle Filter}

\begin{aug}
\author[A]{\fnms{Kari} \snm{Heine}\ead[label=e1,mark]{k.m.p.heine@bath.ac.uk}}\and
\author[A]{\fnms{Daniel} \snm{Burrows}\ead[label=e2,mark]{dwb26@bath.ac.uk}}
\address[A]{Department of Mathematical Sciences, University of Bath, Bath, UK \printead{e1,e2}}
\end{aug}

\begin{abstract}
We consider situations where the applicability of sequential Monte Carlo particle filters is compromised due to the expensive evaluation of the particle weights. To alleviate this problem, we propose a new particle filter algorithm based on the multilevel approach. We show that the resulting multilevel bootstrap particle filter (MLBPF) retains the strong law of large numbers as well as the central limit theorem of classical particle filters under mild conditions. Our numerical experiments demonstrate up to 85\% reduction in computation time compared to the classical bootstrap particle filter, in certain settings. While it should be acknowledged that this reduction is highly application dependent, and a similar gain should not be expected for all applications across the board, we believe that this substantial improvement in certain settings makes MLBPF an important addition to the family of sequential Monte Carlo methods.
\end{abstract}


\end{frontmatter}


\section{Introduction}
\label{sec:introduction}

Sequential Monte Carlo (SMC) methods, or particle filters \cite{gordon_et_al93,doucet_et_al01} are popular computational tools for approximate inference with hidden Markov models (HMM). Particle filter approximation is based on a random sample of weighted particles, where the weights are obtained by evaluating the conditional density of the realised observations, i.e.~the likelihood, for each particle. In certain applications the evaluation of this conditional density may be expensive, and therefore, due to the typically large number of particles, the use of particle filters may become less appealing, if not outright infeasible.

Problems may arise if, for example, the observations are modelled by a system of differential equations with unknown parameters, and these parameters in turn are modelled as a discrete time Markov process representing the hidden signal process. In this case, the parameters are of primary interest while the solution to the differential equations itself is secondary \citep{hong_et_lian12, gelman_et_al96}. More specifically, suppose that a $\Y$-valued observation $Y$ is modelled for a given $\X$-valued hidden state $X$ as
\begin{align*}
Y = h(f_{X}(\ell)) + V,
\end{align*}
where $h: \R^{p}\to \Y$ is some function, $f_{x}: L \to \R^{p}$ is the solution of the above-mentioned system of differential equations on domain $L$, parameterised by $x \in \X$, and $V$ is an additive $\Y$-valued noise term. In this case, to evaluate the weights for particles $x^1,\ldots,x^N \in \X$, for some $N\in \N \defeq \{1,2,\ldots\}$, one would have to find the solution $f_{x^i}$ for each $1 \leq i \leq N$, which in many cases can only be done numerically, thereby making the weight evaluation expensive for large $N$.

Another possibly problematic scenario arises in the context of big (high dimensional) data. Suppose that the observations take values in $\Y = \R^{p}$, where $p$ is large, and that the covariance of $y = (y_{1},\ldots,y_{p})^{T}$ does not admit any specific independence structure. In this case, even a simple observation model 
\begin{align*}
Y = h(X) + V,
\end{align*}
where $V$ denotes additive Gaussian noise with covariance $\Sigma$, would imply that to find the weight for a single particle $x \in \X$ and a given  observation $y$, one would have to evaluate
\begin{align*}
(y - h(x))^T\Sigma^{-1}(y - h(x)),
\end{align*}
requiring $\mathcal{O}(p^2)$ operations in general, which may be too expensive for large $p$.

To overcome the challenges of this kind, we propose a new SMC algorithm based on the principles of multilevel Monte Carlo (MLMC) \citep{heinrich01}. The key idea of MLMC is to introduce approximations at different levels of accuracy and to control the cost-accuracy trade-off of these approximations; computationally inexpensive low level approximations capture the target roughly, while the more expensive and accurate higher level approximations are used for fine tuning the estimates. Computational gain arises, when the expensive high level approximations are used for estimating the error of the lower level approximations, instead of the target quantity directly.
 
MLMC is known to improve the efficiency from classical Monte Carlo in the context of simulating stochastic differential equations \citep{giles08}, but in the context of HMMs, the situation is not equally well understood; the definition of multilevel sequential Monte Carlo is not equally well established, partly due to the ambiguity of what it may mean, and therefore, depending on the meaning, it may not be clear how the results of \citep{giles08} could be applied, or extended, to the sequential context. One example of pioneering work in multilevel sequential Monte Carlo is \citep{beskos_et_al15}, which we will discuss more closely in Section \ref{sec:literature}. We give a rigorous definition of a novel \emph{multilevel bootstrap particle filter} (MLBPF) algorithm as a generalisation of the classical bootstrap particle filter (BPF)~\cite{gordon_et_al93}. Our theoretical analysis shows that MLBPF retains the strong law of large numbers and central limit theorem that are known to hold for the classical BPF~\cite{chopin04, delmoral04, del1999central, crisan_et_doucet02}. We also demonstrate the power of the multilevel approach with two numerical applications.

We wish to emphasise that the classical MLMC literature is focused on models involving stochastic differential equations, but our approach is quite generic and applies to a broad class of problems. We assume only that the trade-off between the cost and accuracy of evaluating the likelihood can be controlled, but we make no assumptions on the mathematical causes of this trade-off, be it numerical solving of differential equations, optimisation problems, complex matrix calculations etc. In the context of differential equations, this trade-off is typically controlled by the mesh size of the solver, but other scenarios are possible, such as the big data example above which we will revisit in the numerical experiments in Section \ref{sec:numerical}.

\subsection{Formal problem statement}

We consider the problem of approximating a generally intractable filter $\widehat{\pi} = (\widehat{\pi}_{n})_{n\geq 0}$ and the associated prediction filter $\pi = (\pi_{n})_{n\geq 0}$ such that
\begin{align*}
\widehat{\pi}_{n} &= \P(X_{n} \in \,\cdot\mid Y_0=y_0,\ldots, Y_{n} = y_{n}), \quad n\geq 0 \\
\pi_{n+1} &= \P(X_{n+1} \in \,\cdot\mid Y_0=y_0,\ldots, Y_{n} = y_{n}),\quad n\geq 0 
\end{align*}
are conditional probabilities with respect to the law of a hidden Markov model with signal process $(X_n)_{n\geq 0}$ and observation process $(Y_n)_{n\geq 0}$. The signal process $(X_n)_{n\geq 0}$ takes values in a measurable space $(\X,\mathcal{X})$ such that
\begin{align*}
X_0 \sim \pi_{0}, \qquad X_n \mid X_{n-1} = x_{n-1} \sim K(x_{n-1},\,\cdot\,),\qquad n>0,
\end{align*}
where $K:\X\times\mathcal{X} \to [0,1]$ is a Markov kernel and $\pi_{0}$ is a probability measure on $\mathcal{X}$.
The observation process $(Y_n)_{n\geq 0}$ takes values in a measurable space $(\Y,\mathcal{Y})$ and satisfies
\begin{align*}
Y_n \mid X_n = x_n \sim G(x_n,\,\cdot\,),\qquad n\geq 0,
\end{align*}
for a probability kernel $G:\X\times\mathcal{Y}\to[0,1]$ which admits a density $g(x,\,\cdot\,)$ with respect to a $\sigma$-finite measure on $\mathcal{Y}$, for all $x \in \X$. For a fixed  realisation $(y_n)_{n\geq 0}$ of observations, we write $g_n(\,\cdot\,) = g(\,\cdot\,,y_n)$, for all $n\geq 0$. 

The proposed MLBPF approximations, which we denote by  $(\widehat{\pi}^{N}_{n})_{n\geq 0}$ and $(\pi^{N}_{n})_{n\geq 0}$, respectively for the filter and the prediction filter, are parametrised by the sample size $N \in \N$ and our main results establish the strong law of large numbers as well a central limit theorem for both $(\pi^{N}_{n})_{n\geq 0}$ and $(\widehat{\pi}^{N}_{n})_{n\geq 0}$ as $N\to\infty$. As a by-product of our analysis, we also obtain a central limit theorem for estimating the normalisation coefficient of the exact filter. However, the estimate is not almost surely positive, and hence inapplicable as such as a marginal likelihood estimate in particle MCMC algorithms~\citep{andrieu_et_al10}, for example.

MLBPF is based on approximating $g_{n}$ with $g_n^{\theta}$, where $\theta \in [0,\theta_{\mathrm{max}}]\subset \R$ denotes the parameter we use for controlling the above-mentioned cost-accuracy trade-off such that for small values of $\theta$, the cost of evaluating $g_n^\theta$ is low but the approximation error is large, and vice versa. Note particularly, that we assume $g_n = g_n^{\theta_{\mathrm{max}}}$, i.e.~for $\theta = \theta_{\mathrm{max}}$ the likelihood is evaluated exactly. While there may be situations where the exact evaluation of $g_n$ is infeasible, we regard $\theta_{\mathrm{max}}$ to represent the highest realistically feasible accuracy of the model, and essentially assume that this is indistinguishable from the exact model.

From now on, we  will simplify the notation by writing $g_{n}^0,\ldots,g_{n}^{L}$ instead of  $g_n^{\theta_0},\ldots,g_n^{\theta_L}$ for the approximations of $g_n$ at different levels.

\subsection{Literature review and the organisation of the paper} 
\label{sec:literature}

MLMC methodology dates back to \citep{heinrich01}, but it has later gained notable popularity in the context of stochastic differential equations \citep{giles08}. It was shown in \citep{giles08}, that under verifiable conditions, the use of multilevel approaches can reduce the order of complexity. The classical MLMC theory is focused on estimating specific integrals with respect to probabiliy measures, rather than the probability measures themselves. Therefore the theory of multilevel Monte Carlo is not immediately applicable to sequential Monte Carlo where the focus is on approximating probability measures by appropriately weighted particles, instead of specific integrals. For this reason, there is relatively little literature on integrating the multilevel methods with sequential Monte Carlo.

An important contribution to multilevel sequential Monte Carlo is \citep{beskos_et_al15}, which demonstrated how so-called SMC samplers \citep{delmoral_et_al06} can be used for computing multilevel approximations. It was also shown how the complexity theorem of \cite{giles08} can be extended to these multilevel SMC (MLSMC) samplers. However, the approach of \citep{beskos_et_al15} is notably different from our work as the sequentiality of the MLSMC samplers arises from an artificial Markov process across the approximation levels{\color{red},} instead of the HMM of the real world system whose state we wish to estimate. Essentially, \citep{beskos_et_al15} uses a standard sequential SMC sampler to generate a multilevel particle approximation of a non-sequential inference problem. In contrast, we propagate (i.e.~resample and mutate) a multilevel particle approximation across filter iterations to approximate a sequential inference problem.  This leads to algorithmic complications that are avoided in \citep{beskos_et_al15} by focusing on non-sequential inference problems only. We show how these complications can be resolved without losing the key asymptotic properties of the classical BPF. This leads to the MLBPF algorithm, which to our knowledge is the first known instance of a multilevel SMC \emph{filter} as opposed to a multilevel SMC \emph{sampler}.

Another recent advance in combining multilevel methods with sequential Monte Carlo is~\cite{prescott_et_baker20}, which considered the context of approximate Bayesian computations (ABC) using SMC sampling~\cite{beaumont_et_al2002,toni_et_al09}. Although their development of the multifidelity ABC-SMC algorithm gives rise to issues similar to those that we encounter in the context of MLBPF, the context is quite different from ours. First, similar to the MLSMC samplers of  \cite{beskos_et_al15}, multifidelity ABC-SMC is not immediately suitable for inference with HMMs. Secondly, ABC aims at avoiding the likelihood evaluations altogether, while our approach is based on evaluating  the likelihood approximately.

The remainder of this paper is organised as follows. In Section \ref{sec:MLBPF} we present our rationale for the multilevel particle filtering and define the MLBPF algorithm. In Sections \ref{sec:SLLN} and \ref{sec:CLT} we prove the strong law of large numbers and central limit theorem for MLBPF, respectively. In Section \ref{sec:numerical} we demonstrate the potential of MLBPF with two different applications and finally in Section \ref{sec:conclusions} we summarise our conclusions from the experiments and theoretical analysis.
 
\section{Multilevel Bootstrap Particle Filter}
\label{sec:MLBPF}

In the exact filter update, $\widehat{\pi}_{n}$ is obtained from $\pi_{n}$ and $g_{n}$ according to (see e.g.~\cite{delmoral04})
\begin{align}\label{eq:exact update}
\widehat{\pi}_{n}(\varphi) = \frac{\pi_{n}(g_{n}\varphi)}{\pi_{n}(g_{n})}, \qquad \varphi \in \boundborel(\X),~n \geq 0,
\end{align}
where $\boundborel(\X)$ denotes the set of bounded and measurable functions defined on $\X$. The key idea of the multilevel methodology is that the integral in the numerator of \eqref{eq:exact update} can be decomposed into a telescoping sum 
\begin{align*}
\pi_{n}(g_{n}\varphi) = \sum_{\ell=0}^{\Lmax} \pi_{n}\left ( \Delta g_{n}^{\ell}\varphi\right ),
\end{align*}
where $\Delta g^{\ell}_{n} = g^{\ell}_{n} - g^{\ell-1}_{n}$ with the convention that $g^{-1}_{n}\equiv 0$ for all $n\geq 0$, and $L+1 \in \N$ is the number of approximation levels. 
After constructing a similar decomposition for the denominator $\pi_{n}(g_{n})$ of \eqref{eq:exact update}, we have
\begin{align}\label{eq:decomposition form}
\widehat{\pi}_{n}(\varphi) &= \sum_{\ell=0}^{\Lmax} p_{n,\ell}\widehat{\pi}_{n,\ell}(\varphi),
\end{align}
where
\begin{align}\label{eq:decomposition components}
p_{n,\ell} = \frac{\pi_{n}(\Delta g^{\ell}_{n} )}{\sum_{\ell=0}^{\Lmax}\pi_{n}(\Delta g^{\ell}_{n})} \qquad \text{and} \qquad \widehat{\pi}_{n,\ell}(\varphi) = \frac{\pi_{n}(\Delta g^{\ell}_{n}\varphi)}{\pi_{n}(\Delta g^{\ell}_{n} )}.
\end{align}
In what follows, we will construct an algorithm for computing particle approximations for the level-specific decomposition components $\widehat{\pi}_{n,\ell}$. This is not a trivial task, as these components of the decomposition \eqref{eq:decomposition form} are in general \emph{signed measures} instead of positive probability measures. 

This observation leads to considerations that are atypical to SMC algorithms because the resulting particle systems may consequently contain particles that have negative weights. This gives rise to complications when rejuvenating the particles by resampling. It is important to keep in mind however, that the emergence of these negative weights is not any kind of undesired artefact of an approximation scheme or lack of numerical accuracy, but a natural property of the \emph{exact} representation of a probability measure $\widehat{\pi}_{n}$ as a linear combination of signed measures, as defined in \eqref{eq:decomposition form} and \eqref{eq:decomposition components}. Therefore, except for the surmountable practical challenges of knowing how to handle the negative weights, there is no obvious downside to them --- quite the opposite: they enable us to leverage the flexibility of the signed measure decomposition \eqref{eq:decomposition form} to enable rigorous corrections of the inexpensive low level filter approximations, resulting in a convergent and computationally more efficient filter algorithm.

\subsection{Algorithm}

For each level $0\leq \ell \leq \Lmax$, we denote the level specific sample size by $N_{\ell} = c_{\ell}N$, where $c_{\ell}$, $N\in \N$, with the assumption that $c_{\Lmax} = 1$. Thus, the total sample size is  $\Ntot(N) \defeq \sum_{\ell=0}^{\Lmax} c_{\ell}N $. By defining $I_{\ell}(N) \defeq \sum_{i=0}^{\ell-1} c_{i}N$ for all all $0 \leq \ell \leq \Lmax$ with the convention that $\sum_{a}^{b}(\,\cdot\,) = 0$ whenever $a>b$, we can partition the particle indices $\{1,\ldots,\Ntot(N)\}$ into subsets 
\begin{align*}
 P_{\ell}^{N} \defeq \left\{I_{\ell}(N)+1,\ldots,I_{\ell+1}(N)\right\},\qquad 0 \leq \ell \leq \Lmax,
\end{align*}
where $P^{N}_{\ell}$ represents the set of particle indices associated with level $\ell$. The multilevel bootstrap particle filter is defined in Algorithm \ref{alg:MLBPF}.

\begin{algorithm}
\begin{algorithmic}
\caption{Multilevel Bootstrap Particle Filter (MLBPF)}\label{alg:MLBPF}
\State \verb+% Initialisation+
\For{$i = 1,\ldots,\Ntot(N)$}
\State $\xi^{i}_{n} \sim \pi_{0}$
\State $w^{i}_{0} = 1$.
\EndFor
\For{$n \geq 0$}
	\State \verb+% Calculate weights for each level+
	\For{$0 \leq \ell \leq \Lmax$}
	\For{$i \in P^{N}_{\ell}$}
	\State $\widetilde{w}_{n}^{i} = N_{\ell}^{-1}(g_{n}^{\ell}(\xi^{i}_{n}) - g_{n}^{\ell-1}(\xi^{i}_{n})){w}^{i}_{n}$
	\EndFor
	\EndFor
	\State \verb+% Signed resampling+
	\For{$i=1,\ldots,\Ntot(N)$}
		\State $\widehat{\xi}^{i}_{n} \sim \dfrac{\sum_{i=1}^{\Ntot(N)}|\widetilde{w}_{n}^{i}|\delta_{\xi^{i}_{n}}}{\sum_{i=1}^{\Ntot(N)}|\widetilde{w}_{n}^{i}|}$ and $\widehat{w}^{i}_{n} = \sign\left (\sum_{i=1}^{\Ntot(N)}\widetilde{w}^{i}_{n}\I[\widehat{\xi}^{i}_{n} = \xi^{i}_{n}]\right )$ 
	\EndFor
	\State \verb+% Mutation+
	\For{$i=1,\ldots,\Ntot(N)$}
		\State $\xi^{i}_{n+1} \sim K(\widehat{\xi}^{i}_{n},\,\cdot\,)$ and $w^{i}_{n+1} = \widehat{w}_{n}^{i}$
	\EndFor
\EndFor
\end{algorithmic}
\end{algorithm}

We see immediately that for $L = 0$, MLBPF reduces to the classical BPF, but for $L>0$ there are notable differences between MLBPF and BPF due to the signed weights $\widetilde{w}^{1}_{n},\ldots,\widetilde{w}^{\Ntot(N)}_{n}$, as there is no guarantee that $g_{n}^{\ell} - g_{n}^{\ell-1}$ is non-negative. Consequently, resampling cannot be carried out in the usual manner by drawing an independent sample of size $\Ntot(N)$ proportionally to the weights $\widetilde{w}^{1}_{n},\ldots,\widetilde{w}^{\Ntot(N)}_{n}$, but instead, the sample is drawn from the distribution proportional to the \emph{total variation measure} $$\sum_{i=1}^{\Ntot(N)}|\widetilde{w}_{n}^{i}|\delta_{\xi^{i}_{n}}.$$
This further implies that in order to keep the approximation convergent to the correct limiting measure, the particles cannot be assigned constant weights after the resampling step as in BPF, but instead, the weights $\widehat{w}^{1}_{n},\ldots,\widehat{w}^{N}_{n}$ must be constant in modulus with varying sign, i.e.
\begin{align*}
\widehat{w}^{i}_{n} = \sign\left (\widetilde{w}_{n}(\widehat{\xi}^{i}_{n})\right ), \qquad n \geq 0,~1 \leq i \leq N, 
\end{align*}
where $\widetilde{w}_{n}: \{\xi^{1}_{n},\ldots,\xi^{\Ntot(N)}_{n}\} \to \R$, is defined as
\begin{align*}
\widetilde{w}_{n}(\xi) \defeq \sum_{i=1}^{\Ntot(N)}\widetilde{w}^{i}_{n}\I[\xi = \xi^{i}_{n}], \qquad \xi \in \{\xi^{1}_{n},\ldots,\xi^{\Ntot(N)}_{n}\},~n\geq 0,
\end{align*}
and $\sign(x) = \I[x>0]-\I[x<0]$, as usual. A natural approximation for $\widehat{\pi}_{n}$ obtained from Algorithm \ref{alg:MLBPF} is then
\begin{align}\label{eq:approximations}
\widehat{\pi}^{N}_{n} = \frac{\sum_{i=1}^{\Ntot(N)}\widehat{w}^{i}_{n}\delta_{\widehat{\xi}^{i}_{n}}}{\sum_{i=1}^{(N)}\widehat{w}^{i}_{n}},\qquad n\geq 0.
\end{align}
\begin{remark}
We consider only the approximation $\widehat{\pi}_{n}^{N}$ in \eqref{eq:approximations} for the filtering distribution $\widehat{\pi}_{n}$. This is the approximation obtained \emph{after the resampling}~\cite{delmoral04}. In general, the approximation \emph{prior to resampling} is more accurate, but for the purposes of proving convergence, the approximation after the resampling is more relevant. Moreover, as the approximation before resampling is better, our results immediately extend to it as well. 
\end{remark}
\begin{remark}
Our main results, Theorem \ref{thm:SLLN main} and Theorem \ref{thm:CLT} below, also hold for the prediction filter approximation 
\begin{align*}
\pi^{N}_{n} = \frac{\sum_{i=1}^{\Ntot(N)} w^{i}_{n}\delta_{\xi^{i}_{n}}}{\sum_{i=1}^{\Ntot(N)}w^{i}_{n}}
\end{align*}
as a trivial by-product of our analysis.
\end{remark}

Before moving on to the theoretical analysis, we conclude this section by imposing the following mild assumptions that are assumed to hold throughout the reminder of this paper:
\begin{assumption}\label{ass:positive kernel}
For all $x \in \X$, $K(x,\,\cdot\,)$ admits a strictly positive density  
with respect to a $\sigma$-finite measure on $\mathcal{X}$.
\end{assumption}
\begin{assumption}
For all $n \in \N$, $g_n > 0$.
\end{assumption}
Moreover, we will use the standard integral operator notation $K(\varphi) = \int \varphi(x)K(\,\cdot\,,\mathrm{d}x)$ for the kernel $K$ throughout the paper.  

\section{Strong Law of Large Numbers}
\label{sec:SLLN}

\newcommand{\positivemeasure}{\mathscr{M}_{+}}
\newcommand{\negativemeasure}{\mathscr{M}_{-}}

\begin{theorem}\label{thm:SLLN main}
For all bounded and measurable $\varphi: \X\to\R$ and all $n\geq 0$
\begin{align*}
\widehat{\pi}^{N}_{n}(\varphi) \asto{N\to\infty} \widehat{\pi}_{n}(\varphi).
\end{align*}
\end{theorem}
Our proof of Theorem \ref{thm:SLLN main} is by induction and
it is based on the asymptotics of the \emph{level-specific unnormalised signed measures}
\begin{align}\label{eq:unnormalised measures}
\gamma^{N}_{n,\ell}(\varphi) \defeq \frac{1}{N_{\ell}}\sum_{i\in P_{\ell}^{N}} w^{i}_{n}\varphi(\xi^{i}_{n}) \quad \text{and} \quad \widehat{\gamma}^{N}_{n,\ell}(\varphi) \defeq \frac{1}{N_{\ell}}\sum_{i\in P_{\ell}^{N}} \widehat{w}^{i}_{n}\varphi(\widehat{\xi}^{i}_{n}) 
\end{align}
for all $n \geq 0$, $N\geq 1$, $0 \leq \ell \leq \Lmax$ and $\varphi\in\boundborel(\X)$. Throughout our analysis, including the proof of the central limit theorem, these measures are regarded as the basic building blocks, which themselves are not of interest, but  from which all other measures of interest, such as $\pi^{N}_n$ and $\widehat{\pi}^{N}_{n}$, can be obtained through some functional mapping. The strategy is to prove the desired asymptotics first for the measures in \eqref{eq:unnormalised measures} and then show that the same results extend to the derived measures as well.

We start by showing (Lemma \ref{lem:a.s. level spec filter limits}) that the almost sure convergence of $\gamma^{N}_{n,\ell}(\varphi)$ implies the almost sure convergence of  $\widehat{\gamma}^{N}_{n,\ell}(\varphi)$, which in turn yields (Lemma \ref{lem:as convergence update})
\begin{align}\label{eq:level spec approximations}
\widehat{\pi}^{N}_{n,\ell}(\varphi) = \frac{\widehat{\gamma}^{N}_{n,\ell}(\varphi)}{\widehat{\gamma}^{N}_{n,\ell}(1)} \asto{N\to\infty} \widehat{\pi}_n(\varphi).
\end{align}
To complete the induction, we also show that the asymptotics we have established for the resampled particles $\widehat{\xi}^{1}_{n},\ldots,\widehat{\xi}^{N}_{n}$ are preserved in the mutation step (Lemma \ref{lem:as convergence mutation}).
Theorem \ref{thm:SLLN main} is then a straightforward corollary of \eqref{eq:level spec approximations}.

Note that in \eqref{eq:level spec approximations}, the weak limit of $\widehat{\pi}^{N}_{n,\ell}$  is $\widehat{\pi}_{n}$, instead of $\widehat{\pi}_{n,\ell}$. To understand this somewhat unexpected result, recall that in the resampling step of Algorithm \ref{alg:MLBPF} we sample
\begin{align*}
\widehat{\xi}^{i}_{n} \sim \dfrac{\sum_{i=1}^{\Ntot(N)}|\widetilde{w}_{n}^{i}|\delta_{\xi^{i}_{n}}}{\sum_{i=1}^{\Ntot(N)}|\widetilde{w}_{n}^{i}|} = \sum_{\ell=0}^{\Lmax}\frac{ \sum_{i\in P^{N}_{\ell}}|\widetilde{w}_{n}^{i}|}{\sum_{\ell=0}^{\Lmax} \sum_{i\in P^{N}_{\ell}}|\widetilde{w}_{n}^{i}|}\frac{\sum_{i\in P^{N}_{\ell}}|\widetilde{w}_{n}^{i}|\delta_{\xi^{i}_{n}}}{\sum_{i\in P^{N}_{\ell}}|\widetilde{w}_{n}^{i}|}, 
\end{align*}
which is analogous to \eqref{eq:decomposition form}. Thus we see that $\widehat{\xi}^{i}_{n}$ is drawn from the marginal total variation measure, obtained by summing over all level-specific total variation measures. Therefore, with appropriate weighting, $(\widehat{\xi}^{i}_{n})_{i \in P^{N}_{\ell}}$ approximates asymptotically $\widehat{\pi}_{n}$ rather than $\widehat{\pi}_{n,\ell}$ which is integrated out by the marginalisation across the levels $0 \leq \ell \leq \Lmax$.

It should also be noted that due to this implicit marginalisation in the resampling step, $(\widehat{\xi}^{i}_{n})_{i \in P^{N}_{\ell}}$ bears no connection to the specific approximation level $\ell$ and the explicit dependency on $\ell$ in the notation should be understood only as an index over $\Lmax+1$ conditionally iid samples. The chosen notation is nevertheless justified as it indicates that at iteration $n+1$, after $(\widehat{\xi}^{i}_{n})_{i \in P^{N}_{\ell}}$ is mutated, the corresponding particles will be weighted using the level-specific likelihood difference $\Delta g_{n+1}^{\ell}$.

From \eqref{eq:level spec approximations} it seems sufficient to focus on the asymptotics of $(\gamma^{N}_{n,\ell})_{n\geq 0}$ only, but due to the resampling according to the total variation measure, we also have to study the asymptotics of the total variation measure $|\gamma^{N}_{n,\ell}|$ of $\gamma^{N}_{n,\ell}$. In general, the weak convergence of $\gamma_{n,\ell}^{N}$ to $\gamma_{n}$ does not imply the weak convergence of $|\gamma_{n,\ell}^{N}|$ to $|\gamma_{n}|$ (see, e.g.~\citep[Corollary 8.4.8]{bogachev07}), and therefore the asymptotics of the total variation measure need to be confirmed separately. 

According to the discussion above, Theorem \ref{thm:SLLN main} now holds by the following asymptotic result for $(\gamma^{N}_{n,\ell})_{n\geq 0}$ and Lemma \ref{lem:as convergence update} below.
\begin{proposition}\label{prop:gamma recursion}
For all $\varphi \in \boundborel(\X)$, $n\geq 0$, and $0 \leq \ell \leq \Lmax$ we have 
\begin{align}\label{eq:main gamma convergence}
\gamma_{n,\ell}^{N}(\varphi) \asto{N\to\infty} \gamma_{n}(\varphi) \quad\text{and}\quad |\gamma_{n,\ell}^{N}|(\varphi) \asto{N\to\infty} \eta_{n}(\varphi),
\end{align}
where 
\begin{align}\label{eq:explicit gamma recursions}
\gamma_{n+1}(\varphi) = \pi_{n+1}(\varphi)\widehat{\gamma}_{n}(1),
\quad\text{and}\quad
\eta_{n+1}(\varphi) = \widehat{\eta}_{n}(K(\varphi))
\end{align}
and
\begin{align}\label{eq:hat limits}
\widehat{\gamma}_{n}(\varphi) = \frac{\sum_{\ell=0}^{\Lmax}\gamma_{n}(\Delta g_{n}^{\ell}\varphi)}{\sum_{\ell=0}^{\Lmax}\eta_{n}(|\Delta g_{n}^{\ell}|)} \quad \text{and} \quad \widehat{\eta}_{n}(\varphi) = \frac{\sum_{\ell=0}^{\Lmax}\eta_{n}(|\Delta g_{n}^{\ell}|\varphi)}{\sum_{\ell=0}^{\Lmax}\eta_{n}(|\Delta g_{n}^{\ell}|)}
\end{align}
and 
 $\gamma_{0} = \eta_{0} = \pi_{0}$.
\end{proposition}

\begin{remark}
From Proposition \ref{prop:gamma recursion} we see that our analysis involves two more measure sequences that are non-standard in SMC literature, namely $(\widehat{\eta}_n)_{n\geq 0}$ and $(\eta_n)_{n\geq 0}$. These are the filter and prediction filter sequences that result when the exact likelihood $g_n$ is replaced with $\sum_{\ell=0}^{L} |\Delta g_n^{\ell}|$. Although this substitution leads to  a well-defined filter, there is no real world counterpart or interpretation to $(\widehat{\eta}_n)_{n\geq 0}$ and $(\eta_n)_{n\geq 0}$. They are purely theoretical constructions, and as we see from \eqref{eq:main gamma convergence}, $\eta_{n}$ is  the weak limit of the level-specific empirical total variation measure approximations $|\gamma^{N}_{n,\ell}|$.
\end{remark}

Proposition \ref{prop:gamma recursion} also admits the following Corollary, which in addition to providing some insight to the relation of $\gamma_n$ and $\eta_n$ will be used in the proof of the central limit theorem. 

\begin{corollary}\label{cor:positivity of difference}
For all $n > 0$, $\eta_{n} - \gamma_{n}$ and $\widehat{\eta}_{n} - \widehat{\gamma}_{n}$ are positive measures.

\end{corollary}

The proofs of Proposition \ref{prop:gamma recursion} and Corollary \ref{cor:positivity of difference} are given in Section \ref{sec:proof for SLLN}. The proof of Proposition \ref{prop:gamma recursion} is by induction, the induction assumption being that Proposition \ref{prop:gamma recursion} holds at rank $n$. The induction step is essentially established by the following three Lemmata, the first of which establishes the asymptotics for the total variation measure after the resampling step at time $n$, under the induction assumption.
\begin{lemma}\label{lem:a.s. level spec filter limits}
If Proposition \ref{prop:gamma recursion} holds for some $n\geq 0$, then 
\begin{align}
\widehat{\gamma}_{n,\ell}^{N}(\varphi) &\asto{N\to\infty} \widehat{\gamma}_{n}(\varphi)\label{eq:as gamma normaliser limit} \\
|\widehat{\gamma}^{N}_{n,\ell}|(\varphi) &\asto{N\to\infty} \widehat{\eta}_{n}(\varphi)\label{eq:as level spec absolute filter limit}
\end{align}
for all $\varphi \in \boundborel(\X)$ and $0 \leq \ell \leq \Lmax$. 
\end{lemma}
\begin{remark}\label{rem:sign partition limits}
If we write $P^{N\pm}_{\ell} = \{i \in P^{N}_{\ell}: \sign(w^{i}_{n}) = \pm 1\},$ then 
\begin{align*} 
\frac{1}{c_\ell N}\sum_{i\in P^{N+}_{\ell}}\varphi(\xi^{i}_{n}) 
 &=\frac{1}{2}\left (|\gamma_{n,\ell}^{N}|(\varphi) + \gamma_{n,\ell}^{N}(\varphi) \right )\asto{N\to\infty} \frac{1}{2}\left (\eta_{n}(\varphi) + \gamma_{n}(\varphi)\right ),
\end{align*}
and similarly
\begin{align*}
\frac{1}{c_\ell N}\sum_{i\in P^{N-}_{\ell}}\varphi(\xi^{i}_{n}) = \frac{1}{2}\left ( |\gamma_{n,\ell}^{N}|(\varphi) - \gamma_{n,\ell}^{N}(\varphi) \right )\asto{N\to\infty} \frac{1}{2}\left (\eta_{n}(\varphi) -\gamma_{n}(\varphi) \right ).
\end{align*}
Moreover, by Corollary \ref{cor:positivity of difference} we see that both limits are strictly positive for a strictly positive $\varphi \in \boundborel(\X)$.
\end{remark}

The second Lemma establishes the asymptotics for the level specific filter approximation at time $n$ under the induction assumption.
\begin{lemma}\label{lem:as convergence update}
If Proposition \ref{prop:gamma recursion} holds for some $n\geq 0$, then
\begin{align*}
\widehat{\pi}^{N}_{n,\ell}(\varphi) - \widehat{\pi}_{n}(\varphi) \asto{N\to\infty} 0.
\end{align*}
for all $\varphi \in \boundborel(\X)$ and $0 \leq \ell \leq \Lmax$. 
\end{lemma}

The last of the three Lemmata does not require the induction assumption as it only involves the asymptotics of the mutation step which by definition is based on simulating from the signal kernel and therefore this result holds irrespective of the aymptotics of the resampled particles $\widehat{\xi}^{1}_{n},\ldots,\widehat{\xi}^{\Ntot(N)}_{n}$.
\begin{lemma}\label{lem:as convergence mutation}
For all $\varphi \in \boundborel(\X)$, $n\geq 0$, and $0\leq \ell \leq \Lmax$ we have
\begin{align}\label{eq:mutation asymptotics part I}
\frac{1}{c_{\ell}N}\sum_{i\in P_{\ell}^{N}}\sign(\widetilde{w}_{n}(\widehat{\xi}^{i}_{n}))(\varphi(\xi^{i}_{n+1})-K(\varphi)(\widehat{\xi}^{i}_{n})) &\asto{N\to\infty} 0.
\end{align}
and 
\begin{align}\label{lem:as convergence mutation part II}
\frac{1}{c_{\ell}N} \sum_{i\in P_{\ell}^{N}} \varphi(\xi^{i}_{n+1}) - \frac{1}{c_{\ell}N}\sum_{i\in P_{\ell}^{N}} K(\varphi)(\widehat{\xi}^{i}_{n}) &\asto{N\to\infty} 0.
\end{align}
\end{lemma}

\subsection{Proofs for the Strong Law of Large Numbers}
\label{sec:proof for SLLN}

This section contains only the proofs for Section \ref{sec:SLLN} and it can be skipped at first reading.

\begin{proof}[\namedlabel{proof:gamma recursion}{Proof of Proposition \ref{prop:gamma recursion}}]
The initialisation of the recursion follows immediately by the definition of $\gamma_{0,\ell}^{N}$, $|\gamma_{0,\ell}^{N}|$ and by repeating the proof of Lemma \ref{lem:as convergence mutation} for \eqref{lem:as convergence mutation part II} where $\xi^{i}_{n+1}$ is replaced with $\xi^{i}_{0}$ and $K(\varphi)(\widehat{\xi}^{i}_{n})$ with $\pi_{0}(\varphi)$. To prove the claim by induction, we assume that \eqref{eq:main gamma convergence} holds at rank $n$ and we show that it also holds at rank $n+1$. 
By definition
\begin{align*}
\gamma^{N}_{n+1,\ell}(\varphi) = \frac{1}{c_{\ell}N}\sum_{i\in P_{\ell}^{N}}w^{i}_{n+1}\varphi(\xi^{i}_{n+1}) = \frac{1}{c_{\ell}N}\sum_{i\in P_{\ell}^{N}}\sign(\widetilde{w}_{n}(\widehat{\xi}^{i}_{n}))\varphi(\xi^{i}_{n+1}),
\end{align*}
and thus by \eqref{eq:mutation asymptotics part I} of Lemma \ref{lem:as convergence mutation}, it suffices to have
\begin{align*}
\frac{1}{c_{\ell}N}\sum_{i\in P^{N}_{\ell}}\sign(\widetilde{w}_{n}(\widehat{\xi}^{i}_{n}))K(\varphi)(\widehat{\xi}^{i}_{n}) \asto{N\to\infty} \pi_{n+1}(\varphi)\frac{ \sum_{\ell=0}^{\Lmax}\gamma_{n}(\Delta g^{\ell}_{n})}{\sum_{\ell=0}^{\Lmax}\eta_{n}(|\Delta g^{\ell}_{n}|)},
\end{align*}
but because $\widehat{\pi}_{n}(K(\varphi)) = \pi_{n+1}(\varphi)$, this follows immediately from \eqref{eq:as gamma normaliser limit} of Lemma \ref{lem:a.s. level spec filter limits} together with 
\begin{align}\label{eq:gamma limit 1}
\widehat{\pi}^{N}_{n,\ell}(K(\varphi)) = \frac{\sum_{i\in P^{N}_{\ell}}\sign(\widetilde{w}_{n}(\widehat{\xi}^{i}_{n}))K(\varphi)(\widehat{\xi}^{i}_{n})}{\sum_{i\in P^{N}_{\ell}}\sign(\widetilde{w}_{n}(\widehat{\xi}^{i}_{n}))} \asto{N\to\infty} \widehat{\pi}_{n}(K(\varphi))
\end{align}
which in turn holds by Lemma \ref{lem:as convergence update}, completing the proof for $(\gamma_{n})_{n\geq 0}$. 

For $(\eta_{n})_{n\geq 0}$ it suffices to observe that, by Lemma \ref{lem:a.s. level spec filter limits}
\begin{align*}
\frac{1}{c_{\ell}N}\sum_{i \in P_{\ell}^{N}}K(\varphi)(\widehat{\xi}^{i}_{n}) \asto{N\to\infty} \frac{\sum_{\ell=0}^{\Lmax}\eta_{n}(|\Delta g^{\ell}_{n}|K(\varphi))}{\sum_{\ell=0}^{\Lmax}\eta_{n}(|\Delta g^{\ell}_{n}|)},
\end{align*}
and thus the claim follows by \eqref{lem:as convergence mutation part II} of Lemma \ref{lem:as convergence mutation}.
\end{proof}

\begin{proof}[\namedlabel{proof:positivity of difference}{Proof of Corollary \ref{cor:positivity of difference}}]
By \eqref{eq:explicit gamma recursions} and \eqref{eq:hat limits},
\begin{align*}
\widehat{\gamma}_n(1) = \frac{\pi_{n}(g_n)}{\sum_{\ell=0}^{\Lmax}\eta_{n}(|\Delta g_{n}^{\ell}|)}\widehat{\gamma}_{n-1}(1) = \prod_{q=0}^{n}\frac{\pi_{q}(g_q)}{\sum_{\ell_{q}=0}^{\Lmax}\eta_{q}(|\Delta g_{q}^{\ell_{q}}|)} \qquad n > 0.
\end{align*}
It is easy to show that 
\begin{align*}
\pi_{n+1}(\varphi) = \frac{\pi_0(g_0K(g_1K(\cdots g_nK(\varphi)\cdots)))}{\prod_{q=0}^{n}\pi_{q}(g_q)},
\end{align*}
and therefore, because $g_n = \sum_{\ell=0}^{L}\Delta g_n^{\ell}$, we can write
\begin{align*}
\gamma_{n+1}(\varphi) = \frac{\pi_0\left (K^{n}\left (\sum_{\ell_0=0}^{L}\cdots \sum_{\ell_n=0}^{L}\prod_{q=0}^{n}\Delta g_q^{\ell_q}K(\varphi)\right )\right )}{\prod_{q=0}^{n}\sum_{\ell_{q}=0}^{\Lmax}\eta_{q}(|\Delta g_{q}^{\ell_{q}}|)},
\end{align*}
where $K^n(\varphi) = K_1(K_2(\cdots K_n(\varphi)\cdots))$ with $K_i = K$ for all $1 \leq i \leq n$, i.e.~$K^n$ is the $n$ fold iterate of integral operator $K$. On the other hand, it follows from \eqref{eq:explicit gamma recursions} and \eqref{eq:hat limits} also that 
\begin{align*}
\eta_{n+1}(\varphi) = \frac{\sum_{\ell=0}^{\Lmax}\eta_{n}(|\Delta g_{n}^{\ell}|K(\varphi))}{\sum_{\ell=0}^{\Lmax}\eta_{n}(|\Delta g_{n}^{\ell}|)} = \frac{\pi_0\left (K^{n}\left (\sum_{\ell_0=0}^{L}\cdots \sum_{\ell_n=0}^{L}\prod_{q=0}^{n}|\Delta g_q^{\ell_{q}}|K(\varphi)\right )\right )}{\prod_{q=0}^{n}\sum_{\ell_{q}=0}^{\Lmax}\eta_{q}(|\Delta g_{q}^{\ell_{q}}|)}.
\end{align*}
Therefore, it suffices to determine the sign of the measure 
\begin{align*}
\pi_0\left (K^{n}\left (\sum_{\ell_0=0}^{L}\cdots \sum_{\ell_n=0}^{L}\prod_{q=0}^{n}\left (|\Delta g_q^{\ell_{q}}|- \Delta g_q^{\ell_{q}}\right )K(\varphi)\right )\right ),
\end{align*}
which is a positive measure due to Assumption \ref{ass:positive kernel}. The positivity of $\widehat{\eta}_{n} - \widehat{\gamma}_{n}$ follows from observing that because $\eta_n-\gamma_n$ and $\gamma_n$ are positive, we have $$\eta_n(|\Delta g^{\ell}_n|\varphi) -\gamma_n(\Delta g^{\ell}_n\varphi) \geq \eta_n(|\Delta g^{\ell}_n|\varphi) - \gamma_n(|\Delta g^{\ell}_n|\varphi) \geq 0$$ for any non-negative $\varphi \in \boundborel(\X)$. Notice that $\eta_n-\gamma_n$ and $\widehat{\eta}_n- \widehat{\gamma}_{n}$ are either strictly positive or zero measures in which case $\eta_n = \gamma_n$ and $\widehat{\eta}_{n}=\widehat{\gamma}_n$.

\end{proof}

\begin{proof}[\namedlabel{proof:as level spec filter limits}{Proof of Lemma \ref{lem:a.s. level spec filter limits}}]

Define for all $0 \leq \ell \leq \Lmax$ and $N \in \N$,  $\widehat{U}^{N}_{0,\ell} = 0$ and
\begin{align*}
\widehat{U}^{N}_{\rho,\ell} = \frac{1}{c_{\ell}\sqrt{N}}\sum_{i\in\calI_{\ell}^{N}(\rho)}\left(\varphi(\widehat{\xi}^{i}_{n}) - \frac{\sum_{j=1}^{\Ntot(N)}|\widetilde{w}^{j}_{n}|\varphi(\xi^{j}_{n})}{\sum_{j=1}^{\Ntot(N)}|\widetilde{w}^{j}_{n}|}\right),~ 1\leq \rho\leq N,~\varphi \in \boundborel(\X),
\end{align*}
where $\calI^{N}_{\ell}(\rho) = \{I_{\ell}(N)+(\rho-1)c_{\ell}+1,\ldots,I_{\ell}(N)+\rho c_{\ell}\}$. We also define the $\sigma$-algebras $(\G^{N}_{\rho})_{0\leq \rho \leq N}$ as 
\begin{align}\label{eq:sigma algebra for resampling}
\G^{N}_{0} = \F^{N}_{n}, \quad \G^{N}_{\rho} = \G^{N}_{\rho-1} \vee \bigvee_{0 \leq \ell \leq \Lmax}\bigvee_{i\in \calI^{N}_{\ell}(\rho)}\sigma\left(\widehat{\xi}^{i}_{n}\right),\quad  1 \leq \rho \leq N,
\end{align}
where $\F^{N}_{n} \subset \F$ is generated by the particles $\xi_{q}^{i}$ and $\widehat{\xi}_{p}^{i}$, where $1\leq i \leq \Ntot(N)$, $0\leq q \leq n$  and $0\leq p < n$.
By the definition of the resampling step in Algorithm \ref{alg:MLBPF}, $\E[\widehat{U}^{N}_{\rho,\ell}\mid \G^{N}_{\rho-1}] = 0$ almost surely. Moreover, $\widehat{U}^{N}_{\rho,\ell}$ is clearly $\G^{N}_{\rho}$-measurable, making $(U^{N}_{\rho,\ell},\,\G^{N}_{\rho})_{0 \leq \rho \leq N,\,N>0}$ a triangular martingale difference array. We also clearly have $|\widehat{U}^{N}_{\rho,\ell}| \leq 2\|\varphi\|/\sqrt{N}$ and therefore, by the Burkholder-Davis-Gundy theorem \citep{burkholder_et_al72}
\begin{align*}
\E\left[\left|\frac{1}{c_{\ell}N}\sum_{i\in P_{\ell}^{N}}\varphi(\widehat{\xi}^{i}_{n}) - \frac{\sum_{j=1}^{\Ntot(N)}|\widetilde{w}^{j}_{n}|\varphi(\xi^{j}_{n})}{\sum_{j=1}^{\Ntot(N)}|\widetilde{w}^{j}_{n}|} \right|^r\,\Bigg|\,\G^{N}_{0}\right] &= \frac{1}{N^{r/2}}\E\left[\left|\sum_{\rho=1}^{N} \widehat{U}^{N}_{\rho,\ell}\right|^{r}\,\Bigg|\,\G^{N}_{0}\right] \\ &\leq \frac{B_{r}2^{r}\|\varphi\|^{r}}{N^{r/2}},
\end{align*}
for some $B_r$ depending only on $r$. Hence by Markov's inequality and the Borel-Cantelli lemma 
\begin{align}\label{eq:interm limit}
\frac{1}{c_{\ell}N}\sum_{i\in P_{\ell}^{N}}\varphi(\widehat{\xi}^{i}_{n}) - \frac{\sum_{i=1}^{\Ntot(N)}|\widetilde{w}^{i}_{n}|\varphi(\xi^{i}_{n})}{\sum_{i=1}^{\Ntot(N)}|\widetilde{w}^{i}_{n}|} \asto{N\to\infty}0.
\end{align}
Under the induction assumption
\begin{align}
\sum_{i=1}^{\Ntot(N)}|\widetilde{w}^{i}_{n}|\varphi(\xi^{i}_{n})
= \sum_{\ell=0}^{\Lmax} |\gamma_{n,\ell}^{N}|(|\Delta g^{\ell}_{n}|\varphi)&\asto{N\to\infty} \sum_{\ell=0}^{\Lmax} \eta_{n}(|\Delta g^{\ell}_{n}|\varphi), \label{eq:absolute limit}\\
\sum_{i=1}^{\Ntot(N)}\widetilde{w}^{i}_{n}\varphi(\xi^{i}_{n}) = \sum_{\ell=0}^{\Lmax} \gamma_{n,\ell}^{N}(\Delta g^{\ell}_{n}\varphi) &\asto{N\to\infty} \sum_{\ell=0}^{\Lmax} \gamma_{n}(\Delta g^{\ell}_{n}\varphi).\label{eq:non-absolute limit}
\end{align}
Therefore \eqref{eq:as level spec absolute filter limit} follows from  \eqref{eq:absolute limit} and \eqref{eq:interm limit} and \eqref{eq:as gamma normaliser limit} follows from replacing $\varphi(\widehat{\xi}^{i}_{n})$ with $\sign(\widetilde{w}_{n}(\widehat{\xi}^{i}_{n}))\varphi(\widehat{\xi}^{i}_{n})$, observing that $|\widetilde{w}^{i}_{n}|\sign(\widetilde{w}_{n}(\widehat{\xi}^{i}_{n})) = \widetilde{w}^{i}_{n}$, and finally by using \eqref{eq:non-absolute limit} and \eqref{eq:interm limit}.
\end{proof}

\begin{proof}[\namedlabel{proof:as convergence update}{Proof of Lemma \ref{lem:as convergence update}}]
Let $(\G^{N}_{\rho})_{0\leq \rho \leq \Ntot}$ be as defined in \eqref{eq:sigma algebra for resampling} in the proof of Lemma \ref{lem:a.s. level spec filter limits} and for all $0 \leq \ell \leq \Lmax$, we define $U^{N}_{0,\ell} = 0$ and 
\begin{align}\label{eq:martingale diff}
U^{N}_{\rho,\ell} = \frac{1}{c_{\ell}\sqrt{ N}}\sum_{i\in\calI^{N}_{\ell}(\rho)}\sign(\widetilde{w}_{n}(\widehat{\xi}^{i}_{n}))\left(\varphi(\widehat{\xi}^{i}_{n}) - \frac{\sum_{j=1}^{\Ntot(N)}\widetilde{w}^{j}_{n}\varphi(\xi_{n}^{j})}{\sum_{j=1}^{\Ntot(N)}\widetilde{w}^{j}_{n}}\right),
\end{align}
for all $1 \leq \rho \leq N$ and some $\varphi \in \boundborel(\X)$. Clearly $U^{N}_{\rho,\ell}$ is $\G^{N}_{\rho}$-measurable and
\begin{align}
c_{\ell}\sqrt{N}\E[U^{N}_{\rho,\ell} \mid \G^{N}_{\rho-1}] 
&= \sum_{i\in\calI^{N}_{\ell}(\rho)}\E\Bigg[\sign(\widetilde{w}_{n}(\widehat{\xi}^{i}_{n}))\left(\varphi(\widehat{\xi}^{i}_{n}) - \frac{\sum_{j=1}^{\Ntot(N)}\widetilde{w}^{j}_{n}\varphi(\xi_{n}^{j})}{\sum_{j=1}^{\Ntot(N)}\widetilde{w}^{j}_{n}}\right)\Bigg|\,\G^{N}_{\rho-1}\Bigg] \nonumber\\
&=\frac{c_{\ell}}{\sum_{j=1}^{\Ntot(N)}|\widetilde{w}^{j}_{n}|}\sum_{i=1}^{\Ntot(N)}|\widetilde{w}^{i}_{n}|\sign(\widetilde{w}^{i}_{n})\left(\varphi(\xi^{i}_{n}) - \frac{\sum_{j=1}^{\Ntot(N)}\widetilde{w}^{j}_{n}\varphi(\xi_{n}^{j})}{\sum_{j=1}^{\Ntot(N)}\widetilde{w}^{j}_{n}}\right) \nonumber\\
&=\frac{c_{\ell}}{\sum_{j=1}^{\Ntot(N)}|\widetilde{w}^{j}_{n}|}\sum_{i=1}^{\Ntot(N)}\widetilde{w}^{i}_{n}\left(\varphi(\xi^{i}_{n}) - \frac{\sum_{j=1}^{\Ntot(N)}\widetilde{w}^{j}_{n}\varphi(\xi_{n}^{j})}{\sum_{j=1}^{\Ntot(N)}\widetilde{w}^{j}_{n}}\right) \nonumber\\
&=\frac{c_{\ell}}{\sum_{j=1}^{\Ntot(N)}|\widetilde{w}^{j}_{n}|}\left(\sum_{i=1}^{\Ntot(N)}\widetilde{w}^{i}_{n}\varphi(\xi^{i}_{n}) - \sum_{i=1}^{\Ntot(N)}\widetilde{w}^{i}_{n}\frac{\sum_{j=1}^{\Ntot(N)}\widetilde{w}^{j}_{n}\varphi(\xi_{n}^{j})}{\sum_{j=1}^{\Ntot(N)}\widetilde{w}^{j}_{n}}\right)= 0, \label{eq:zero expectation}
\end{align}
almost surely and so $(U^{N}_{\rho,\ell},\,\G^{N}_{\rho})_{0 \leq \rho \leq N,\,N\geq 0}$ is a triangular martingale difference array. By \eqref{eq:non-absolute limit} and \eqref{eq:explicit gamma recursions} 
\begin{align}\label{eq:as filter limit}
\frac{\sum_{j=1}^{\Ntot(N)}\widetilde{w}^{j}_{n}\varphi(\xi^{j}_{n})}{\sum_{j=1}^{\Ntot(N)}\widetilde{w}^{j}_{n}} \asto{N\to\infty} \frac{\sum_{\ell=0}^{L}\gamma_{n}(\Delta g_{n}^{\ell}\varphi)}{\sum_{\ell=0}^{L}\gamma_{n}(\Delta g_{n}^{\ell})} = \frac{\pi_{n}(g_{n}\varphi)}{\pi_{n}(g_{n})},
\end{align}
and therefore, for any $\delta > 0$,  there exists almost surely $N_{\delta}\in \N$ such that for all $N>N_{\delta}$
\begin{align*}
\left|\frac{\sum_{i=1}^{\Ntot(N)}\widetilde{w}^{i}_{n}\varphi(\xi^{i}_{n})}{\sum_{i=1}^{\Ntot(N)}\widetilde{w}^{i}_{n}}\right| < \left|\frac{\pi_{n}(g_{n}\varphi)}{\pi_{n}(g_{n})}\right| + \delta.
\end{align*}
From the definition \eqref{eq:martingale diff} we see that for all $N>N_{\delta}$,
\begin{align}\label{eq:upper bound for the martingale difference}
|U^{N}_{\rho,\ell}| \leq  \frac{C_{\varphi}}{\sqrt{N}},\quad\text{where}\quad C_{\varphi} \defeq \|\varphi\| + \left|\frac{\pi_{n}(g_{n}\varphi)}{\pi_{n}(g_{n})}\right| + \delta,
\end{align}
and therefore, by the Burkhold-Davis-Gundy theorem
\begin{align}\label{eq:bgd conclusion}
\frac{1}{N^{r/2}}\E\left[\left|\sum_{\rho=1}^{N} U^{N}_{\rho,\ell}\right|^{r}\,\Bigg|\, \G^{N}_{0}\right] \leq \frac{B_{r}C_{\varphi}^{r}}{N^{r/2}},
\end{align}
for some $B_r$ depending only on $r$. We also observe that
\begin{align}
\frac{1}{\sqrt{N}}\sum_{\rho=1}^{N}U^{N}_{\rho,\ell} 
&=\frac{1}{c_{\ell}N}\sum_{\rho=1}^{N}\sum_{i\in\calI^{N}_{\ell}(\rho)}\sign(\widetilde{w}_{n}(\widehat{\xi}_{n}^{i}))\left(\varphi(\widehat{\xi}^{i}_{n})-\frac{\sum_{j=1}^{\Ntot(N)}\widetilde{w}^{j}_{n}\varphi(\xi^{j}_{n})}{\sum_{j=1}^{\Ntot(N)}\widetilde{w}^{j}_{n}}\right) \nonumber\\
&=\frac{1}{c_{\ell}N}\sum_{i\in P_{\ell}^{N}}\sign(\widetilde{w}_{n}(\widehat{\xi}_{n}^{i}))\varphi(\widehat{\xi}^{i}_{n}) - \frac{1}{c_{\ell}N}\sum_{i\in P_{\ell}^{N}}\sign(\widetilde{w}_{n}(\widehat{\xi}_{n}^{i}))\frac{\sum_{j=1}^{\Ntot(N)}\widetilde{w}^{j}_{n}\varphi(\xi^{j}_{n})}{\sum_{j=1}^{\Ntot(N)}\widetilde{w}^{j}_{n}} \nonumber\\
&=\Bigg(\frac{1}{c_{\ell}N}\sum_{i\in P_{\ell}^{N}}\sign(\widetilde{w}_{n}(\widehat{\xi}_{n}^{i}))\Bigg)\Bigg(\frac{\sum_{i\in P_{\ell}^{N}}\sign(\widetilde{w}_{n}(\widehat{\xi}_{n}^{i}))\varphi(\widehat{\xi}^{i}_{n})}{\sum_{i\in P_{\ell}^{N}}\sign(\widetilde{w}_{n}(\widehat{\xi}_{n}^{i}))} - \frac{\sum_{j=1}^{\Ntot(N)}\widetilde{w}^{j}_{n}\varphi(\xi^{j}_{n})}{\sum_{j=1}^{\Ntot(N)}\widetilde{w}^{j}_{n}}\Bigg),
\nonumber
\end{align}
which, by noting that,
\begin{align*}
\widehat{\pi}^{N}_{n,\ell}(\varphi) = \frac{\sum_{i\in P_{\ell}^{N}}\sign(\widetilde{w}_{n}(\widehat{\xi}_{n}^{i}))\varphi(\widehat{\xi}^{i}_{n})}{\sum_{i\in P_{\ell}^{N}}\sign(\widetilde{w}_{n}(\widehat{\xi}_{n}^{i}))}
\end{align*}
enables the decomposition
\begin{align*}
\widehat{\pi}^{N}_{n,\ell}(\varphi) - \widehat{\pi}_{n}(\varphi) =  \frac{\frac{1}{\sqrt{N}}\sum_{\rho=1}^{N}U^{N}_{\rho,\ell}}{\frac{1}{N_{\ell}}\sum_{i\in P_{\ell}^{N}}\sign(\widetilde{w}_{n}(\widehat{\xi}^{i}_{n}))} + \frac{\sum_{j=1}^{\Ntot(N)}\widetilde{w}^{j}_{n}\varphi(\xi^{j}_{n})}{\sum_{j=1}^{\Ntot(N)}\widetilde{w}^{j}_{n}} - \frac{\pi_{n}(g_{n}\varphi)}{\pi_{n}(g_{n})}.
\end{align*}
By \eqref{eq:bgd conclusion}, Markov's inequality and the Borel-Cantelli lemma we have
\begin{align*}
\frac{1}{\sqrt{N}}\sum_{\rho=1}^{N}U^{N}_{\rho,\ell} \asto{N\to\infty} 0
\end{align*}
and therefore the claim follows from the limit \eqref{eq:as gamma normaliser limit} of Lemma \ref{lem:a.s. level spec filter limits} for $\varphi = 1$, which is strictly positive, and \eqref{eq:as filter limit}.
\end{proof}

\begin{proof}[\namedlabel{proof:as convergence mutation}{Proof of Lemma \ref{lem:as convergence mutation}}]
For all $0 \leq \ell \leq \Lmax$ and $N\in\N$, we define $\widetilde{U}^{N}_{0,\ell} = 0$ and
\begin{align}\label{eq:mutation martingale 1}
\widetilde{U}^{N}_{\rho,\ell} = \frac{1}{c_{\ell}\sqrt{N}}\sum_{i\in\calI_{\ell}^{N}(\rho)}\sign(\widetilde{w}_{n}(\widehat{\xi}^{i}_{n}))(\varphi(\xi^{i}_{n+1})-K(\varphi)(\widehat{\xi}^{i}_{n})),
\end{align}
for all $1 \leq \rho \leq N$ and some $\varphi \in \boundborel(\X)$. Moreover we define $\sigma$-algebras $(\widetilde{\G}^{N}_{\rho})_{0 \leq \rho \leq N,\,N>0}$, such that
\begin{align*}
\widetilde{\G}^{N}_{0} = \widehat{\F}^{N}_{n},\quad \widetilde{\G}^{N}_{\rho} = \widetilde{\G}^{N}_{\rho-1} \vee \bigvee_{0 \leq \ell \leq \Lmax}\bigvee_{i\in \calI_{\ell}^{N}(\rho)}\sigma\left(\xi^{i}_{n+1}\right),\quad 1 \leq \rho \leq N.
\end{align*}
where $\widehat{\F}^{N}_{n} \subset \F$ is generated by the particles $\xi_{q}^{i}$ and $\widehat{\xi}_{q}^{i}$, where $0\leq q \leq n$ and $1\leq i \leq \Ntot(N)$. By the definition in Algorithm \ref{alg:MLBPF},  $\xi^{i}_{n+1} \sim K(\widehat{\xi}^{i}_{n},\,\cdot\,)$ where $\xi^{1}_{n+1},\ldots,\xi^{\Ntot(N)}_{n+1}$ are conditionally independent given $\widetilde{\G}^{N}_{0}$. Therefore, we have $\E[\widetilde{U}^{N}_{\rho,\ell}\mid \widetilde{\G}^{N}_{\rho-1}]=0$ almost surely for all $1 \leq \rho \leq N$ and we also see that $\widetilde{U}^{N}_{\rho,\ell}$ is $\widetilde{\G}^{N}_{\rho}$-measurable, making $(\widetilde{U}^{N}_{\rho,\ell},\,\widetilde{\G}^{N}_{\rho})_{0\leq \rho \leq N,\,N>0}$ a triangular martingale difference array. We also have $|\widetilde{U}^{N}_{\rho,\ell}| \leq 2\|\varphi\|/\sqrt{N}$, and thus by the Burkholder-Davis-Gundy theorem
\begin{align*}
\E\left[\left|\frac{1}{c_{\ell}N}\sum_{i\in P_{\ell}^{N}}\sign(\widetilde{w}_{n}(\widehat{\xi}^{i}_{n}))(\varphi(\xi^{i}_{n+1})-K(\varphi)(\widehat{\xi}^{i}_{n}))\right|^r\,\Bigg|\, \widetilde{\G}^{N}_{0}\right] \leq \frac{B_{r}2^{r}\|\varphi\|^{r}}{{N}^{r/2}},
\end{align*}
for some $B_r$ depending only on $r$. Hence by Markov's inequality and the Borel-Cantelli lemma we have \eqref{eq:mutation asymptotics part I}. Assertion \eqref{lem:as convergence mutation part II} follows by repeating the above proof but with $\sign(\widetilde{w}_{n}(\widetilde{\xi}^{i}_{n}))$ omitted from the definition of $\widetilde{U}^{N}_{\rho,\ell}$.
\end{proof}

\section{Central Limit Theorem}
\label{sec:CLT}

\begin{theorem}\label{thm:CLT}
For all bounded and measurable $\varphi: \X\to\R$ and all $n\geq 0$,
\begin{align*}
\sqrt{N}\left (\widehat{\pi}^{N}_{n}(\varphi) - \widehat{\pi}_{n}(\varphi)\right ) &\indist{N\to\infty} \calN\left (0, \widehat{\sigma}^{2}_{n}(\varphi)\right ),
\end{align*}
for some $\widehat{\sigma}^{2}_{n}(\varphi) \in (0,\infty)$. 
\end{theorem}

Central limit theorems, analogous to Theorem \ref{thm:CLT} above,  can be found in the literature for various SMC algorithms~\cite{chopin04,heine_et_al20,del1999central,delmoral04,kunsch05}. 
A key difference between our proof of Theorem \ref{thm:CLT} for MLBPF and the proofs found in the literature arises from the interpretation whereby we see $\widehat{\pi}^{N}_{n}$ (resp.~${\pi}^{N}_{n}$) as the result of a specific functional mapping being applied to the level-specific measures in the collection $\Gamma_{n,n}$ (resp.~$\Gamma_{n,n-1}$) which we formally define as
\begin{align*}
\Gamma_{n,m} = 
\left \{\gamma^{N}_{p,\ell},~\widehat{\gamma}^{N}_{q,\ell},~|\gamma^{N}_{p,\ell}|,~|\widehat{\gamma}^{N}_{q,\ell}|: 0 \leq \ell \leq \Lmax,~0 \leq p \leq n,~0 \leq q \leq m\right \},\quad n\in\N,~m \in \{n-1,n\}.
\end{align*}
This means that Theorem \ref{thm:CLT} can be proved by first ensuring the \emph{joint} asymptotic normality for these measures that are the building blocks for the actual measures of interest, i.e.~$\widehat{\pi}^{N}_{n}$ (and ${\pi}^{N}_{n}$). Theorem \ref{thm:CLT} then follows straightforwardly by the $\delta$-method (see e.g.~\cite{delmoral04}). 

To formally state what we mean by the joint asymptotic normality, fix $n \in \N$, $m\in \{n-1,n\}$, and $\Lmax > 0$, set $d=2(\Lmax+1)(n+m+2)$, and define for all $\vt = (t_1,\ldots, t_{d})^T \in \R^{d}$ and all $\vvarphi = (\varphi_1,\ldots,\varphi_d)^T \in \boundborel(\X)^d$
\begin{align}
\begin{split}
\Psi^{N}_{n,m}(\vt,\vvarphi) 
&=\sum_{\ell=0}^{\Lmax}\left (\sum_{p=0}^{n}\left (t^{(1)}_{p,\ell}\left (\gamma^{N}_{p,\ell}(\varphi^{(1)}_{p,\ell}) - \gamma_{p}(\varphi^{(1)}_{p,\ell})\right ) + t^{(2)}_{p,\ell}\left (|\gamma^{N}_{p,\ell}|(\varphi^{(2)}_{p,\ell}) - \eta_{p}(\varphi^{(2)}_{p,\ell})\right )\right )\right .\\
&\qquad\left .+\sum_{q=0}^{m}\left (t^{(3)}_{q,\ell}\left (\widehat{\gamma}^{N}_{q,\ell}({\varphi}^{(3)}_{q,\ell}) - \widehat{\gamma}_{q}({\varphi}^{(3)}_{q,\ell})
\right )+ 
t^{(4)}_{q,\ell}\left (|\widehat{\gamma}^{N}_{q,\ell}|({\varphi}^{(4)}_{q,\ell}) - \widehat{\eta}_{q}({\varphi}^{(4)}_{q,\ell})\right )\right )\right ),
\end{split}
\label{eq:joint asymptotics}
\end{align}
where we have written $t^{(k)}_{p,\ell} = t_{\beta(k,p,l)}$,  $\varphi^{(k)}_{p,\ell} = \varphi_{\beta(k,p,l)}$, and $\beta:(k,p,\ell) \mapsto (4p + (k-1))(\Lmax + 1) + \ell + 1$ simply converts the three dimensional indexing over $k$, $p$ and $\ell$ into a one dimensional index over the set $\{1,\ldots,d\}$. This three dimensional indexing is used for convenience, as it explicitly identifies the filter iteration ($p \in \{0,\ldots,n\}$), approximation level ($\ell \in \{0,\ldots,\Lmax\}$), and the type of the measure ($k \in \{1,2,3,4\}$) corresponding to  one of the four types of measures $\gamma^{N}_{p,\ell}$, $|\gamma^{N}_{p,\ell}|$, $\widehat{\gamma}^{N}_{p,\ell}$ or  $|\widehat{\gamma}^{N}_{p,\ell}|$ as illustrated in \eqref{eq:joint asymptotics}. 
We say that the measures in $\Gamma_{n,m}$ satisfy the joint asymptotic normality if for all $\vt \in \R^{d}$ and all $\vvarphi \in \boundborel(\X)^d$
\begin{align}\label{eq:equiv joint asymptotics}
\sqrt{N}\Psi^{N}_{n,m}(\vt,\vvarphi) \indist{N\to\infty} \calN(0,\vt^T\bGamma_{n,m}(\vvarphi)\vt),
\end{align}
for some symmetric positive semi-definite matrix $\bGamma_{n,m}(\vvarphi)$ of size $d\times d$. By Cram\'er-Wold theorem (see e.g.~\citep{billingsley95}) this then implies that the $d$ individual differences in \eqref{eq:joint asymptotics} are jointly asymptotically normal and Theorem \ref{thm:CLT} follows by $\delta$-method.

The structure of our proof is similar to the proofs found in the literature, see e.g.~\cite{chopin04,kunsch05}; we show that the required joint asymptotic normality holds at initialisation and that it is preserved in the update and mutation steps from which Theorem \ref{thm:CLT} follows by induction. The differences to the existing literature arise primarily from the triangular martingale difference array constructions that are specific to the measures in $\Gamma_{n,m}$ which, in turn, are specific to MLBPF, especially the total variation measures.

Let us emphasise that by \eqref{eq:joint asymptotics}, we consider the joint asymptotic normality not only across the approximation levels, but also over filter iterations. For the proof of Theorem \ref{thm:CLT} this additional complexity would be superfluous, but it enables us to prove results across multiple filter iterations, such as the following central limit theorem for the normalisation term of the filter recursions:
\begin{theorem}
There exists $\sigma_{Z}^{2} > 0$ such that
\begin{align*}
\sqrt{N}\left (\prod_{p=0}^{n-1}\pi^{N}_{p}(g_p) - \E\left [\prod_{p=0}^{n-1}g_p(X_p)\right ]\right ) \indist{N\to\infty} \calN(0,\sigma_{Z}^{2}).
\end{align*}
\end{theorem}

The proof of Theorem \ref{thm:CLT} uses the following well-known auxiliary result, which is hard to find in the literature as a standalone result, and therefore its proof is included for completeness.
\begin{lemma}\label{lem:conditional joint CLT}
Let $(A_{N})_{N > 0}$ and $(B_{N})_{N > 0}$ be sequences of $\X$ valued random variables, such that for all $N \in \N$, $B_{N}$ is $\G_{N}$-measurable, 
\begin{align}\label{eq:joint cond clt primise 1}
\sqrt{N}B_{N} \indist{N\to\infty} B \sim \calN\left(0,\sigma^2_{B}\right),
\end{align}
and 
\begin{align}\label{eq:joint cond clt primise 2}
\E\left[\exp\left(iu\sqrt{N}A_{N}\right) \Big|\, \G_{N} \right] \inprob{N\to\infty} \exp\left(-\frac{u^2}{2}\sigma^{2}_{A}\right).
\end{align}
Then
\begin{align*}
\sqrt{N}\left(A_{N} + B_{N}\right) \indist{N\to\infty} \calN\left(0,\sigma^{2}_{A} + \sigma^{2}_{B}\right).
\end{align*}
\end{lemma}

\subsection{Proofs for the Central Limit Theorem}
\label{sec:proof for CLT}

This section contains only the proofs for Section \ref{sec:CLT} and it can be skipped at first reading.

\begin{proof}[\namedlabel{proof:new update step}{Proof of Theorem \ref{thm:CLT}}]

The proof is by induction and we start with the update step. Make the induction assumption that \eqref{eq:equiv joint asymptotics} holds for some $n \in \N$ and $m=n-1$,
write 
\begin{align*}
\widehat{\vt}_{n} &= (t^{(3)}_{n,0},\ldots,t^{(3)}_{n,\Lmax},t^{(4)}_{n,0},\ldots,t^{(4)}_{n,\Lmax})^T \in \R^{2(\Lmax+1)},\\
\widehat{\vvarphi}_{n} &= (\varphi^{(3)}_{n,0},\ldots,\varphi^{(3)}_{n,\Lmax},\varphi^{(4)}_{n,0},\ldots,\varphi^{(4)}_{n,\Lmax})^T \in \boundborel(\X)^{2(\Lmax + 1)},
\end{align*}
and consider the triangular martingale difference array $(\widehat{U}^{N}_{\rho},\G^{N}_{\rho})_{0\leq \rho \leq N,\,N> 0}$, where
\begin{align*}
\widehat{U}^{N}_{\rho} = \sum_{\ell=0}^{\Lmax}\left (t^{(3)}_{n,\ell}\widehat{U}^{N}_{\rho,\ell}(\ovarphi_{n,\ell}^{(3)}) + t^{(4)}_{n,\ell}\widehat{U}^{N}_{\rho,\ell}(\varphi_{n,\ell}^{(4)})\right ), 
\end{align*}
with
\begin{align*}
\ovarphi_{n,\ell}^{(3)}(\xi^{i}_{n}) = \sign(\widetilde{w}_{n}(\xi^{i}_{n}))\varphi^{(3)}_{n,\ell}(\xi^{i}_{n}) \qquad 1 \leq i \leq \Ntot(N),
\end{align*}
and $\widehat{U}^{N}_{n,\ell}$ is as defined in the proof of Lemma \ref{lem:a.s. level spec filter limits}, except that now we include the dependency on the test function explicitly in the notation. Also note that we have two types of test functions, $\ovarphi_{n,\ell}^{(3)}$ and $\varphi_{n,\ell}^{(4)}$. 

Clearly, by the proof of Lemma \ref{lem:a.s. level spec filter limits},
\begin{align*}
\left |\widehat{U}^{N}_{\rho}\right | \leq \frac{C_{\overline{\vvarphi}}}{\sqrt{N}}\quad\text{where}\quad C_{\overline{\vvarphi}}=  \sum_{\ell=0}^{\Lmax} \left (t^{(3)}_{n,\ell}\|\varphi^{(3)}_{n,\ell}\| + t^{(4)}_{n,\ell}\|\varphi^{(4)}_{n,\ell}\|\right ),
\end{align*}
and hence
\begin{align}
\sum_{\rho=1}^{N}\E\left[\left(\widehat{U}^{N}_{\rho}\right)^{2}\I\left[\left|\widehat{U}^{N}_{\rho}\right|\geq \epsilon\right]\mid \G^{N}_{\rho-1}\right] 
&\leq\frac{C_{\overline{\vvarphi}}^2}{N}\sum_{\rho=1}^{N}\P\Big[\left|\widehat{U}^{N}_{\rho,\ell}\right|\geq \epsilon\mid \G^{N}_{\rho-1}\Big] \nonumber\\
&\leq \frac{C_{\overline{\vvarphi}}^2}{N}\sum_{\rho=1}^{N}\I\Big[\frac{C_{\overline{\vvarphi}}}{\sqrt{N}}\geq\epsilon\Big] \asto{N\to\infty} 0.\label{eq:in prob to zero}
\end{align}
As the levels are conditionally independent given $\G^{N}_{\rho-1}$, and for all $0\leq \ell \leq \Lmax$ we have $\E[\widehat{U}^{N}_{\rho,\ell}(\ovarphi^{(3)}_{n,\ell})\mid \G^{N}_{\rho-1}] = \E[\widehat{U}^{N}_{\rho,\ell}(\varphi^{(4)}_{n,\ell})\mid \G^{N}_{\rho-1}] = 0$, the second moments satisfy
\begin{align*}
\E\left [ \left (\widehat{U}^{N}_{\rho}\right )^{2} \Big|\,\G^{N}_{\rho-1}\right ] &=  \sum_{\ell=0}^{\Lmax}\left [\left (t^{(3)}_{n,\ell}\right )^{2}\E\left [ \left (\widehat{U}^{N}_{\rho,\ell}(\ovarphi^{(3)}_{n,\ell})\right )^{2} \bigg|\,\G^{N}_{\rho-1}\right ] + \left (t^{(4)}_{n,\ell}\right )^{2}\E\left [ \left (\widehat{U}^{N}_{\rho,\ell}(\varphi^{(4)}_{n,\ell})\right )^{2} \bigg|\,\G^{N}_{\rho-1}\right ]\right ] \\
&+2\sum_{\ell=0}^{\Lmax}\sum_{\ell'=\ell+1}^{\Lmax}t^{(3)}_{n,\ell}t^{(4)}_{n,\ell'} \E\left [\widehat{U}^{N}_{\rho,\ell}(\ovarphi^{(3)}_{n,\ell})\widehat{U}^{N}_{\rho,\ell}(\varphi^{(4)}_{n,\ell})\bigg|\,\G^{N}_{\rho-1}\right ].
\end{align*}
By using \eqref{eq:absolute limit}  and \eqref{eq:non-absolute limit}, we can easily find limits
\begin{align*}
\sum_{\rho=1}^{N}\E\left [ \left (\widehat{U}^{N}_{\rho,\ell}(\ovarphi^{(3)}_{n,\ell})\right )^{2} \bigg|\,\G^{N}_{\rho-1}\right ] 
&\asto{N\to\infty} \frac{1}{c_{\ell}}\left ( \widehat{\eta}_{n}((\varphi^{(3)}_{n,\ell})^{2}) - 
\widehat{\gamma}_{n}(\varphi^{(3)}_{n,\ell})^{2}
\right) \\
\sum_{\rho=1}^{N}\E\left [ \left (\widehat{U}^{N}_{\rho,\ell}(\varphi^{(4)}_{n,\ell})\right )^{2} \bigg|\,\G^{N}_{\rho-1}\right ] &\asto{N\to\infty} \frac{1}{c_{\ell}}\left (
\widehat{\eta}_{n}((\varphi^{(4)}_{n,\ell})^{2})-\widehat{\eta}_{n}(\varphi^{(4)}_{n,\ell})^{2}
\right) \\
\sum_{\rho=1}^{N}\E\left [ \widehat{U}^{N}_{\rho,\ell}(\ovarphi^{(3)}_{n,\ell}) \widehat{U}^{N}_{\rho,\ell}(\varphi^{(4)}_{n,\ell})\bigg|\,\G^{N}_{\rho-1}\right ]
&\asto{N\to\infty} \frac{1}{c_{\ell}}
\left (
\widehat{\gamma}_{n}(\varphi^{(3)}_{n,\ell}\varphi^{(4)}_{n,\ell})
- \widehat{\gamma}_{n}(\varphi^{(3)}_{n,\ell})
\widehat{\eta}_{n}(\varphi^{(4)}_{n,\ell})\right ),
\end{align*}
and so
\begin{align}\label{eq:2nd moment limit}
\sum_{\rho=1}^{N} \E\left [ \left (\widehat{U}^{N}_{\rho}\right )^{2} \Big|\,\G^{N}_{\rho-1}\right ] 
\asto{N\to\infty} \widehat{\vt}^{T}_n\bGamma'_{n}(\widehat{\vvarphi}_n)\widehat{\vt}_n,\quad\text{where}\quad\bGamma'_{n}(\widehat{\vvarphi}_n) = \begin{pmatrix}
\mathbf{A} & \mathbf{B} \\ 
\mathbf{B} & \mathbf{C}
\end{pmatrix},
\end{align}
and
\begin{align*}
\mathbf{A} &= \displaystyle\mydiag_{0\leq \ell \leq \Lmax}\frac{1}{c_{\ell}}\left ( \widehat{\eta}_{n}((\varphi^{(3)}_{n,\ell})^{2})- \widehat{\gamma}_{n}(\varphi^{(3)}_{n,\ell})^{2}\right) \\
\mathbf{B} &= \mydiag_{0 \leq \ell \leq \Lmax}\frac{1}{c_{\ell}}
\left (
\widehat{\gamma}_{n}(\varphi^{(3)}_{n,\ell}\varphi^{(4)}_{n,\ell})
- \widehat{\gamma}_{n}(\varphi^{(3)}_{n,\ell})
\widehat{\eta}_{n}(\varphi^{(4)}_{n,\ell})\right ) \\
\mathbf{C} &= \mydiag_{0\leq \ell \leq \Lmax}\frac{1}{c_{\ell}}\left (
\widehat{\eta}_{n}((\varphi^{(4)}_{n,\ell})^{2})-\widehat{\eta}_{n}(\varphi^{(4)}_{n,\ell})^{2}
\right)
,
\end{align*}
where we use the notation $\mydiag_{0\leq \ell \leq \Lmax} a_{\ell} = \mathrm{diag}(a_0,\ldots,a_{L})$. Notice that the limit in \eqref{eq:2nd moment limit} is strictly positive by Corollary \ref{cor:positivity of difference}.

To complete the proof for the update step, consider the decomposition
\begin{align*}
\Psi^{N}_{n,n}(\vt,\vvarphi) 
&=\sum_{\ell=0}^{\Lmax}t^{(3)}_{n,\ell}\left (\widehat{\gamma}^{N}_{n,\ell}(\varphi^{(3)}_{n,\ell})
- \widehat{\gamma}_{n}(\varphi^{(3)}_{n,\ell})\right ) +  
\sum_{\ell=0}^{\Lmax}t^{(4)}_{n,\ell}\left (|\widehat{\gamma}^{N}_{n,\ell}|(\varphi^{(4)}_{n,\ell})
- \widehat{\eta}_{n}(\varphi^{(4)}_{n,\ell})\right ) \\
&\qquad + \Psi^{N}_{n,n-1}(\widehat{\vt}_{0,n},\widehat{\vvarphi}_{0,n}) \\ &= \widehat{A}^{N} + \widehat{B}^{N}
\end{align*}
where $\widehat{\vt}_{0,n}$ and $\widehat{\vvarphi}_{0,n}$ are such that 
$\vt = (\widehat{\vt}^{T}_{0,n},\widehat{\vt}^T_n)^T$, and $\vvarphi = (\widehat{\vvarphi}^{T}_{0,n},\widehat{\vvarphi}^T_n)^T$. Now
\begin{align*}
\widehat{A}^{N} &= \sum_{\ell=0}^{\Lmax}\left [t^{(3)}_{n,\ell}\!\left (\widehat{\gamma}^{N}_{n,\ell}(\varphi^{(3)}_{n,\ell}) - \frac{\sum_{j=1}^{\Ntot(N)}\widetilde{w}_{n}^{j}\varphi^{(3)}_{n,\ell}(\xi^{j}_{n})}{\sum_{j=1}^{\Ntot(N)}|\widetilde{w}_{n}^{j}|}\right ) + t^{(4)}_{n,\ell}\!\left (|\widehat{\gamma}^{N}_{n,\ell}|(\varphi^{(4)}_{n,\ell}) - \frac{\sum_{j=1}^{\Ntot(N)}|\widetilde{w}_{n}^{j}|\varphi^{(4)}_{n,\ell}(\xi^{j}_{n})}{\sum_{j=1}^{\Ntot(N)}|\widetilde{w}_{n}^{j}|}\right )\right ] \\
\widehat{B}^{N} &= \sum_{\ell=0}^{\Lmax}\left [t^{(3)}_{n,\ell}\!\left (\frac{\sum_{j=1}^{\Ntot(N)}\widetilde{w}_{n}^{j}\varphi^{(3)}_{n,\ell}(\xi^{j}_{n})}{\sum_{j=1}^{\Ntot(N)}|\widetilde{w}_{n}^{j}|} - \widehat{\gamma}_{n}(\varphi^{(3)}_{n,\ell})\right)  + t^{(4)}_{n,\ell}\!\left (\frac{\sum_{j=1}^{\Ntot(N)}|\widetilde{w}_{n}^{j}|\varphi^{(4)}_{n,\ell}(\xi^{j}_{n})}{\sum_{j=1}^{\Ntot(N)}|\widetilde{w}_{n}^{j}|} - \widehat{\eta}_{n}(\varphi^{(4)}_{n,\ell})\right)\right ] \\
&+ \Psi^{N}_{n,n-1}(\widehat{\vt}_{0,n},\widehat{\vvarphi}_{0,n}).
\end{align*}
By \eqref{eq:in prob to zero}, \eqref{eq:2nd moment limit} and \citep[Theorem A.3]{douc_et_moulines08}, which we have included in the Appendix as Theorem \ref{thm:douc_moulines} for completeness, we have
\begin{align}\label{eq:douc et moulines part}
\E \left [ \exp\left (iu \sqrt{N}\widehat{A}^{N}\right ) \bigg|\, \G^{N}_{0}\right ] \inprob{N\to\infty} \exp\left (-\frac{u^2}{2}\widehat{\vt}_n^{T}\bGamma'_{n}(\widehat{\vvarphi}_n)\widehat{\vt}_n\right ).
\end{align}
Moreover, by the induction assumption that \eqref{eq:equiv joint asymptotics} holds for $n$ and $m=n-1$, we can apply the $\delta$-method to obtain
\begin{align}\label{eq:delta method part}
\sqrt{N}\widehat{B}^{N} \indist{N\to\infty} \calN\left (0, \vt^{T}\bGamma'_{n,n-1}(\vvarphi)\vt\right ),
\end{align}
for some $\bGamma'_{n,n-1}(\vvarphi) \in \R^{4(\Lmax+1)(n+1) \times 4(\Lmax+1)(n+1)}$, for which a more explicit expression could be found by using the $\delta$-method. Moreover, by Lemma \ref{lem:conditional joint CLT}, the claim that \eqref{eq:equiv joint asymptotics} holds for $n$ and $m=n$  follows from \eqref{eq:douc et moulines part} and \eqref{eq:delta method part}, as we have
\begin{align*}
\sqrt{N}\Psi^{N}_{n,n}(\vt,\vvarphi) \indist{N\to\infty} \calN\left (0,\vt^T\left ( 
\begin{bmatrix} \boldsymbol{0} & \boldsymbol{0} \\
\boldsymbol{0} & \bGamma'_{n}(\widehat{\vvarphi}_{n})\end{bmatrix} + \bGamma'_{n,n-1}(\vvarphi)\right )\vt \right ).
\end{align*}

Next we show that the joint asymptotic normality is preserved by the mutation step. The induction assumption in this case is that \eqref{eq:equiv joint asymptotics} holds for some $n\in \N$ and $m=n$. Consider the decomposition
\begin{align*}
\Psi^{N}_{n+1,n}(\vt,\vvarphi) & =\sum_{\ell=0}^{\Lmax}t^{(1)}_{n+1,\ell}\left (\gamma^{N}_{n+1,\ell}(\varphi^{(1)}_{n+1,\ell}) - \gamma_{n+1}(\varphi^{(1)}_{n+1,\ell})\right )\\&\qquad + \sum_{\ell=0}^{\Lmax}t^{(2)}_{n+1,\ell}\left (|\gamma^{N}_{n+1,\ell}|(\varphi^{(2)}_{n+1,\ell}) - \eta_{n+1}(\varphi^{(2)}_{n+1,\ell})\right )
+ \Psi^{N}_{n,n}(\vt_{0,n},\vvarphi_{0,n}) \\&= A^{N}+B^{N},
\end{align*}
where $\vt_{0,n}$ and $\vvarphi_{0,n}$ are such that $\vt = (\vt_{0,n}^T,\vt_{n+1}^T)^T$ and $\vvarphi = (\vvarphi_{0,n}^T,\vvarphi_{n+1}^T)^T$, where
\begin{align*}
\vt_{n+1} &= (t^{(1)}_{n+1,0},\ldots,t^{(1)}_{n+1,\Lmax},t^{(2)}_{n+1,0},\ldots,t^{(2)}_{n+1,\Lmax})^T \in \R^{2(\Lmax +1)} \\
\vvarphi_{n+1} &= (\varphi^{(1)}_{n+1,0},\ldots,\varphi^{(1)}_{n+1,\Lmax},\varphi^{(2)}_{n+1,0},\ldots,\varphi^{(2)}_{n+1,\Lmax})^T \in \boundborel(\X)^{2(\Lmax +1)},
\end{align*}
and 
\begin{align*}
A^{N} &= \sum_{\ell=0}^{\Lmax} t^{(1)}_{n+1,\ell}\left (\gamma^{N}_{n+1,\ell}(\varphi^{(1)}_{n+1,\ell})  - \widehat{\gamma}^{N}_{n,\ell}(K(\varphi^{(1)}_{n+1,\ell}))\right )\\&\qquad + \sum_{\ell=0}^{\Lmax} t^{(2)}_{n+1,\ell}\left (|\gamma^{N}_{n+1,\ell}|(\varphi^{(2)}_{n+1,\ell}) - 
|\widehat{\gamma}^{N}_{n,\ell}|(K(\varphi^{(2)}_{n+1,\ell}))\right) \\
&= \sum_{\ell=0}^{\Lmax}\sum_{i\in P^{N}_{\ell}}\left (\frac{t_{n+1,\ell}^{(1)}}{c_{\ell}N}w^{i}_{n+1}\left (\varphi^{(1)}_{n+1,\ell}(\xi^{i}_{n+1}) - K(\varphi^{(1)}_{n+1,\ell})(\widehat{\xi}^{i}_{n}) \right ) \right. \\&\qquad \left.+\frac{t_{n+1,\ell}^{(2)}}{c_{\ell}N}\left (\varphi^{(2)}_{n+1,\ell}(\xi^{i}_{n+1}) - K(\varphi^{(2)}_{n+1,\ell})(\widehat{\xi}^{i}_{n}) \right )\right ) \\
B^{N} &= \sum_{\ell=0}^{\Lmax} t^{(1)}_{n+1,\ell}\left (\widehat{\gamma}^{N}_{n,\ell}(K(\varphi^{(1)}_{n+1,\ell})) - \gamma_{n+1}(\varphi^{(1)}_{n+1,\ell})\right) \\&\qquad  + \sum_{\ell=0}^{\Lmax} t^{(2)}_{n+1,\ell}\left (|\widehat{\gamma}^{N}_{n,\ell}|(K(\varphi^{(2)}_{n+1,\ell})) -\eta_{n+1}(\varphi^{(2)}_{n+1,\ell}) \right)
+ \Psi^{N}_{n,n}(\vt_{0,n},\vvarphi_{0,n}).
\end{align*}
For all $0 \leq \ell < \Lmax$, $N \in \N$, and  $i \in P^{N}_{\ell}$, we define $Z_{i,\ell} = w^{i}_{n+1}X_{i,\ell}+Y_{i,\ell}$,
where
\begin{align*}
X_{i,\ell} = \frac{t_{n+1,\ell}^{(1)}}{c_{\ell}}\left (\varphi^{(1)}_{n+1,\ell} - K(\varphi^{(1)}_{n+1,\ell})(\widehat{\xi}^{i}_{n}) \right )~~\text{and}~~
Y_{i,\ell} = \frac{t_{n+1,\ell}^{(2)}}{c_{\ell}}\left (\varphi^{(2)}_{n+1,\ell} - K(\varphi^{(2)}_{n+1,\ell})(\widehat{\xi}^{i}_{n}) \right ).
\end{align*}
Moreover, for all $N\in\N$, we define $\widetilde{U}^{N}_{0} = 0$ and
\begin{align*}
\widetilde{U}^{N}_{\rho} = \frac{1}{\sqrt{N}}\sum_{\ell=0}^{\Lmax}\sum_{i\in \calI^{N}_{\ell}(\rho)}Z_{i,\ell}(\xi^{i}_{n+1}),\qquad 1 \leq \rho \leq N,
\end{align*}
in which case
\begin{align*}
\sqrt{N}A_{N} = \sum_{\rho=1}^{N} \widetilde{U}^{N}_{\rho}.
\end{align*}
Clearly, $\widetilde{U}^{N}_{\rho}$ is $\widetilde{\G}^{N}_{\rho}$-measurable and $\E[\widetilde{U}^{N}_{\rho}\mid \widetilde{\G}^{N}_{\rho-1}]=0$ almost surely, so $(\widetilde{U}^{N}_{\rho},\widetilde{\G}_{\rho})_{0 \leq \rho \leq N,N>0}$ is a triangular martingale difference array. Moreover, 
\begin{align*}
|\widetilde{U}^{N}_{\rho}| \leq \frac{C'_{\vvarphi}}{\sqrt{N}} \quad\text{where} \quad C'_{\vvarphi}= 2\sum_{\ell = 0}^{\Lmax}\left ( t^{(1)}_{n+1,\ell}\|\varphi^{(1)}_{n+1,\ell}\| + t^{(2)}_{n+1,\ell}\|\varphi^{(2)}_{n+1,\ell}\|\right ),
\end{align*}
so similarly to \eqref{eq:in prob to zero}, we have 
\begin{align}\label{eq:douc_moulines premise 1}
\sum_{\rho=1}^{N}\E\left [\left (\widetilde{U}^N_{\rho}\right )^2 \I\left [\left |\widetilde{U}^N_{\rho}\right |\geq \epsilon\right ]\mid \widetilde{G}^{N}_{\rho-1}\right ] &\inprob{N\to\infty} 0.
\end{align}
Now, because for all $0 \leq \ell \leq \Lmax$, $1 \leq \rho \leq N$ and $i \in \calI^{N}_{\ell}(\rho)$, $Z_{i,\ell}$ are conditionally independent given $\widetilde{\G}^{N}_{\rho-1}$ and $\E[Z_{i,\ell} \mid \widetilde{\G}^{N}_{\rho-1}]=0$, we have
\begin{align}\label{eq:douc_moulines premise 2 interm}
\sum_{\rho=1}^{N}\E\left [\left (\widetilde{U}^N_{\rho}\right )^2 \mid \widetilde{\G}^{N}_{\rho-1}\right ] &= \frac{1}{N}\sum_{\ell=0}^{\Lmax}\sum_{i\in P^{N}_{\ell}}K (Z_{i,\ell}^{2}) (\widehat{\xi}^{i}_{n}).
\end{align}
A direct calculation shows that
\begin{align*}
K (Z_{i,\ell}^{2}) (\widehat{\xi}^{i}_{n}) = K((X_{i,\ell} + \sign(\widehat{w}^{i}_n) Y_{i,\ell} )^2)(\widehat{\xi}^{i}_{n}),
\end{align*}
which is clearly non-negative, but in order to determine when it is strictly positive, we need to consider two cases. Case $1^{\circ}$: $\varphi^{(i)}_{n+1,\ell}$ is almost surely a constant for all $i\in\{1,2\}$, i.e.~for all $i\in\{1,2\}$ and some $a_i \in \R$, we have $\varphi^{(i)}_{n+1,\ell} = a_i$ almost surely with respect to the dominating $\sigma$-finite measure of Assumption \ref{ass:positive kernel}, call this measure $\lambda$. In this case $Z^{2}_{i,\ell} = 0$ almost surely for all $0\leq \ell \leq \Lmax$ and $i \in P^{N}_{\ell}$. Case $2^\circ$ (the complement of case $1^\circ$): we write for some $\varepsilon >0$, 
\begin{align*}
E^{\pm}_{\varepsilon} = \left \{x \in \X: K\left (\left (X_{i,\ell} \pm Y_{i,\ell} \right )^2\right )(x) > \varepsilon\right \}.
\end{align*} 
By Assumption \ref{ass:positive kernel}, and the fact that at least one of the functions $\varphi^{(1)}_{n+1,\ell}$ or $\varphi^{(2)}_{n+1,\ell}$ must not be a constant ($\lambda$-a.s.), we know that at either $\eta_n(E^{+}_{\varepsilon})+\gamma_n(E^{+}_{\varepsilon})>0$ or $\eta_n(E^{-}_{\varepsilon})-\gamma_n(E^{-}_{\varepsilon})>0$, or both, for sufficiently small $\varepsilon > 0$. Hence
\begin{align*}
\frac{1}{ N} \sum_{i\in P^{N}_{\ell}} K(Z_{\ell,i}^{2})(\widehat{\xi}^{i}_{n})
&= \frac{1}{ N}\sum_{i \in P^{N+}_{\ell}}K((X_{i,\ell} + Y_{i,\ell} )^2)(\widehat{\xi}^{i}_{n}) + \frac{1}{ N}\sum_{i \in P^{N-}_{\ell}}K((X_{i,\ell} - Y_{i,\ell} )^2)(\widehat{\xi}^{i}_{n}) \nonumber\\
&>\frac{\varepsilon}{N}\left (\sum_{i \in P^{N+}_{\ell}}\I[\widehat{\xi}^{i}_{n} \in E^{+}_{\varepsilon}] + \sum_{i \in P^{N-}_{\ell}}\I[\widehat{\xi}^{i}_{n} \in E^{-}_{\varepsilon}] \right ) \nonumber\\
&\asto{N\to\infty}\frac{\varepsilon c_{\ell}}{2}\left (\eta_n(E_{+})+\gamma_n(E_{+}) + \eta_n(E_{-})-\gamma_n(E_{-})\right ), 
\end{align*}
where the limit is strictly positive by Remark \ref{rem:sign partition limits}. This implies that we have a strictly positive limit
\begin{align*}
&\frac{1}{N} \sum_{i\in P^{N}_{\ell}} K(Z_{\ell,i}^{2})(\widehat{\xi}^{i}_{n}) \asto{N\to\infty} c_\ell\varsigma^{2}_{\ell} > 0, 
\end{align*}
where it is easy to check that
\begin{align*}
&\varsigma^{2}_\ell = \left (\frac{t^{(1)}_{n+1,\ell}}{c_{\ell}}\right )^{2}\widehat{\eta}_{n}(K((\varphi^{(1)}_{n+1,\ell}-K(\varphi^{(1)}_{n+1,\ell}))^{2})) 
+ \left (\frac{t^{(2)}_{n+1,\ell}}{c_{\ell}}\right )^{2}\widehat{\eta}_{n}(K((\varphi^{(2)}_{n+1,\ell}-K(\varphi^{(2)}_{n+1,\ell}))^{2})) \\
&\qquad + \frac{t^{(1)}_{n+1,\ell}t^{(2)}_{n+1,\ell}}{c_{\ell}^2}\widehat{\gamma}_{n}(K((\varphi^{(1)}_{n+1,\ell}-K(\varphi^{(1)}_{n+1,\ell}))(\varphi^{(2)}_{n+1,\ell}-K(\varphi^{(2)}_{n+1,\ell})))).
\end{align*}
From this we conclude by \eqref{eq:douc_moulines premise 2 interm} that
\begin{align*}
\sum_{\rho=1}^{N}\E\left [\left (\widetilde{U}^N_{\rho}\right )^2 \mid \widetilde{\G}^{N}_{\rho-1}\right] \asto{N\to\infty} \sum_{\ell=0}^{\Lmax} c_\ell\varsigma^{2}_{\ell} = \vt^T_{n+1}\bGamma''_{n+1}(\vvarphi_{n+1})\vt_{n+1}\quad \text{where}\quad \bGamma''_{n+1}(\vvarphi_{n+1}) = \begin{pmatrix}\mathbf{A} & \mathbf{B}\\
\mathbf{B} & \mathbf{C}\end{pmatrix},
\end{align*}
and 
\begin{align*}
\mathbf{A} &= \mydiag_{0\leq \ell \leq \Lmax} \frac{1}{c_{\ell}}\widehat{\eta}_{n}(K((\varphi^{(1)}_{n+1,\ell}-K(\varphi^{(1)}_{n+1,\ell}))^{2}))\\
\mathbf{B} &= \mydiag_{0\leq\ell\leq\Lmax}\frac{1}{c_{\ell}}\widehat{\gamma}_{n}(K((\varphi^{(1)}_{n+1,\ell}-K(\varphi^{(1)}_{n+1,\ell}))(\varphi^{(2)}_{n+1,\ell}-K(\varphi^{(2)}_{n+1,\ell})))) \\
\mathbf{C} &= \mydiag_{0\leq \ell \leq \Lmax} \frac{1}{c_{\ell}}\widehat{\eta}_{n}(K((\varphi^{(2)}_{n+1,\ell}-K(\varphi^{(2)}_{n+1,\ell}))^{2})).
\end{align*}
In conclusion, provided that $\vt^T_{n+1}\bGamma''_{n+1}(\vvarphi_{n+1})\vt_{n+1}>0$, which happens if Case $2^\circ$ holds for any $0\leq \ell \leq \Lmax$, Theorem \ref{thm:douc_moulines} in the Appendix states that
\begin{align}\label{eq:douc_moulines conclusion}
\E\left [\exp\left (iu\sqrt{N}A_{N}\right )\mid \widetilde{\G}_{0}^{N}\right] \inprob{N\to\infty} \exp\left (-\frac{u^2}{2}\vt^{T}_{n+1}\bGamma''_{n+1}(\vvarphi_{n+1})\vt_{n+1}\right ),
\end{align}
and finally, by noting that
\begin{align*}
\gamma_{n+1}(\varphi^{(1)}_{n+1,\ell}) = \widehat{\gamma}_{n}(K(\varphi^{(1)}_{n+1,\ell}))\quad \text{and}\quad \eta_{n+1}(\varphi^{(2)}_{n+1,\ell}) = \widehat{\eta}_{n}(K(\varphi^{(2)}_{n+1,\ell})),
\end{align*}
we have by the induction assumption that \eqref{eq:equiv joint asymptotics} holds for $n$ and $m=n$, and the $\delta$-method
\begin{align}\label{eq:consequence of update}
\sqrt{N}B^{N} \indist{N\to\infty} \calN\left (0,\vt^{T}\bGamma_{n,n}''(\vvarphi)\vt\right ),
\end{align}
for some symmetric positive semi-definite $\bGamma_{n,n}''(\vvarphi)$. The claim that \eqref{eq:equiv joint asymptotics} holds for $n+1$ and $m=n$ then follows from \eqref{eq:douc_moulines conclusion}, \eqref{eq:consequence of update} and Lemma \ref{lem:conditional joint CLT}, as we have shown that
\begin{align*}
\sqrt{N}\Psi^{N}_{n+1,n}(\vt,\vvarphi) \indist{N\to\infty} \calN\left (0,\vt^T\left ( \begin{bmatrix}\mathbf{0} & \mathbf{0} \\ \mathbf{0} & \bGamma''_{n}(\vvarphi) \end{bmatrix} + \bGamma''_{n,n}(\vvarphi)  \right )\vt \right ).
\end{align*}
As we know that $\vt^T_{n+1}\bGamma''_{n+1}(\vvarphi_{n+1})\vt_{n+1}$ is non-negative, the only case we still need to consider is $\vt^T_{n+1}\bGamma''_{n+1}(\vvarphi_{n+1})\vt_{n+1}=0$ which occurs only if for all $0\leq \ell \leq \Lmax$ we have Case $1^{\circ}$ above. In this degenerate case  $A_N = 0$ almost surely and the claim follows immediately with a degenerate limiting distribution with zero variance.

Our final task is to initialise the induction, i.e.~we need to show that \eqref{eq:equiv joint asymptotics} holds for $n=0$ and $m=-1$. For this we observe that $\gamma_{0,\ell} = |\gamma_{0,\ell}|$ almost surely and $\gamma_{0}=\eta_{0} = \pi_{0}$. An analysis analogous to the proof of the update step earlier yields 
\begin{align*}
\sqrt{N}\Psi^{N}_{0,-1}(\vt,\vvarphi) \indist{N\to\infty} \calN\left (0, \vt^T\begin{pmatrix}\mathbf{A} & \mathbf{B} \\ \mathbf{B} & \mathbf{C}\end{pmatrix}\vt\right ),
\end{align*}
where $\vt = (t^{(1)}_{0},\ldots,t^{(1)}_{\Lmax},t^{(2)}_{0},\ldots,t^{(2)}_{\Lmax})$, $\vvarphi = (\varphi^{(1)}_{0},\ldots,\varphi^{(1)}_{\Lmax},\varphi^{(2)}_{0},\ldots,\varphi^{(2)}_{\Lmax})$, $\Psi^{N}_{0,-1}$ is as defined in \eqref{eq:joint asymptotics}, with the sum over $q$ being equal to zero, and
\begin{align*}
\mathbf{A} &= \mydiag_{0\leq\ell\leq\Lmax} \frac{1}{c_{\ell}}\pi_{0}((\varphi^{(1)}_{\ell}-\pi_{0}(\varphi^{(1)}_{\ell})^{2}) \\
\mathbf{B} &= \mydiag_{0\leq\ell\leq\Lmax} \frac{1}{c_{\ell}}\pi_{0}((\varphi^{(1)}_{\ell}-\pi_{0}(\varphi^{(1)}_{\ell})(\varphi^{(2)}_{\ell}-\pi_{0}(\varphi^{(2)}_{\ell}))\\
\mathbf{C} &= \mydiag_{0\leq\ell\leq\Lmax} \frac{1}{c_{\ell}}\pi_{0}((\varphi^{(2)}_{\ell}-\pi_{0}(\varphi^{(2)}_{\ell})^{2}).
\end{align*}
The proof is then completed by \eqref{eq:level spec approximations}, the fact that
\begin{align*}
\frac{\gamma_{n}(\varphi)}{\gamma_{n}(1)} = \pi_{n}(\varphi)\quad\text{and}\quad 
\frac{\widehat{\gamma}_{n}(\varphi)}{\widehat{\gamma}_{n}(1)} = \widehat{\pi}_{n}(\varphi),
\end{align*}
and an application of the $\delta$-method. 
\end{proof}

\begin{proof}[\namedlabel{proof:conditional joint CLT}{Proof of Lemma \ref{lem:conditional joint CLT}}]
By the continuous mapping theorem, Slutsky's theorem (see e.g.~ \cite{karr93}), \eqref{eq:joint cond clt primise 1}, and \eqref{eq:joint cond clt primise 2}, 
\begin{align*}
\E\left[\exp\left(iu\sqrt{N}A_{N}\right) \Big|\, \G_{N} \right]\exp\left(iu\sqrt{N}B_{N}\right) \indist{N\to\infty} \exp\left(-\frac{u^2}{2}\sigma^{2}_{A} + iu B\right).
\end{align*}
As the complex exponential is continuous and bounded, we can extend the convergence to the expectations, implying that
\begin{align*}
\lim_{N\to\infty}\E\left[\E\left[\exp\left(iu\sqrt{N}A_{N}\right) \Big|\, \G_{N} \right]\exp\left(iu\sqrt{N}B_{N}\right)\right]
&= \lim_{N\to\infty}\E\left[\exp\left(iu\sqrt{N}A_{N}+iu\sqrt{N}B_{N}\right)\right] \\&= \exp\left(-\frac{u^2}{2}\left(\sigma^{2}_{A}+\sigma^{2}_{B}\right)\right)
\end{align*}
where we have also use the $\G_{N}$-measurability of $B_{N}$. The claim now follows from L\'evy's continuity theorem.
\end{proof}


\section{Numerical results}
\label{sec:numerical}

In this section we consider two applications to demonstrate the performance of MLBPF in improving computational efficiency. The source codes, written in \texttt{C}, are available at \url{https://github.com/heinekmp/MLBPF}. In the experiments of Section \ref{sec:ebb} we used the LAPACK \texttt{dgbsv} solver implemented in the Apple Accelerate framework \cite{accelerate} to solve the ordinary differential equations.

\subsection{Big data}
\label{sec:big data}

Consider the model introduced in Section \ref{sec:introduction} for high dimensional data with a Gaussian AR(1) signal model such that
\begin{align}\label{eq:signal model}
X_0 \sim \mathcal{N}(0,\sigma^2)\quad \text{and}\quad X_n \mid X_{n-1} = x_{n-1} \sim \mathcal{N}(x_{n-1},\sigma^2), \qquad n > 0,
\end{align}
where $\sigma = 0.1$. The measurement function $h$ is assumed to be the identity mapping, and the full $N_{\mathrm{obs}}$ by $N_{\mathrm{obs}}$ observation covariance matrix $\Sigma^{(1)}$ is generated randomly as
\begin{align*}
\Sigma^{(1)}_{i,j} = B_{i,j}\exp(-2|i-j|),
\end{align*}
where $B = AA^T$ and $A_{i,j}$ are independent uniform random numbers on the interval $[0,1)$ for all $i,j \in \{1,\ldots,N_{\mathrm{obs}}\}$. 

A two level MLBPF is considered with level 0 approximation obtained by using simply the diagonal $$\Sigma^{(0)} = \mathrm{diag}\left (\Sigma^{(1)}_{1,1},\ldots,\Sigma^{(1)}_{N_{\mathrm{obs}},N_{\mathrm{obs}}}\right ),$$ of $\Sigma^{(1)}$, i.e.~at level 0, the observations are assumed to be independent. This yields a notable computational saving compared to using the full matrix $\Sigma^{(1)}$ (see \cite{rebeschini_et_handel15} for an alternative approach to similar problem). We set $N_{\mathrm{obs}}=500$ and ran the filter for 50 iterations. The error is measured in terms of mean squared error (MSE) to the exact filter mean, which in this case can be found exactly and efficiently with Kalman filter~\cite{kalman60}. MSE is computed over 50 filter time steps.

It should be pointed out that in addition to approximating the likelihood by taking the diagonal, it turns out that we need to introduce an additional heuristic to improve the performance of MLBPF. We use the approximation $\widetilde{g}^{0}_{n} = Cg^{0}_{n}$, where $g^{0}_{n}$ is the approximation obtained by using the diagonal and 
\begin{align*}
C = \argmin_{c \in \R}\sum_{i\in P^{N}_{1}}\left (cg^{0}_n(\xi_n^{i}) - g^{1}_{n}(\xi^i_{n})\right)^2 = \frac{\sum_{i\in P^{N}_{1}}g^{0}_n(\xi_n^{i})g^{1}_{n}(\xi^i_{n})}{\sum_{i\in P^{N}_{1}}(g^{0}_n(\xi_n^{i}))^2}.
\end{align*}
\begin{remark}
Essentially the additional correction above only scales the approximation $g^{0}_{n}$. In the classical BPF, this scaling would be cancelled due to the normalisation of the weights, but with MLBPF, the level 0 and level 1 weights are normalised jointly, and hence this scaling does affect the performance.
\end{remark}

Note that although we confine ourselves to the use of diagonal approximation $\Sigma^{(0)}$ only, in general it would be possible to consider $\Sigma^{(0)}$ to be a band matrix with varying band width. This would naturally lead to multilevel approximations instead of the two level approximation considered here. Due to the level 0 accuracy being fixed to that of the diagonal approximation, we have only one degree of freedom to optimise the MLBPF performance: the sample allocation $(N_0,N_1)$ for the two levels. We found empirically different sample allocations, reported in Table \ref{tab:time matched sample sizes}, that result in a computation time close to that of a BPF with sample size $N=250$, call it BPF1. Note that although we do not have to assume the sequence $c_{0},\ldots,c_{L}$ to be decreasing, in practice this will be the case as fewer particles should be allocated to the higher and more expensive levels. 
\begin{table}[b]
\caption{Level specific sample sizes $(N_0,N_1)$ for MLBPF matching the time complexity of BPF with $N=250$.}
\label{tab:time matched sample sizes}
\begin{tabular}{c|cccccccccc}
$N_0$ & 68000 & 60656 & 53312 & 45968 & 38624 & 31008 & 23664 & 16320 & 8976 & 1360 \\
$N_1$ & 0 & 27 & 54 & 81 & 108 & 136 & 163 & 190 & 217 & 245 \\
RMSE $(10^{-2})$ & 3.079 & 5.434 & 4.259 & 2.263 & 2.212 & 1.972 & 1.621 & 2.079 & 2.629 & 34.690
\end{tabular}
\end{table}
We also ran another BPF with larger sample size $N=1750$, call it BPF2. This sample size was empirically determined to produce approximately the same level of MSE as the MLBPF with the best choice of $(N_0,N_1)$. The computation time of BPF2 was approximately 7 times that of BPF1, or the MLBPF.

\begin{figure}[t]
\begin{tikzpicture}
\node at (0,0) {\includegraphics{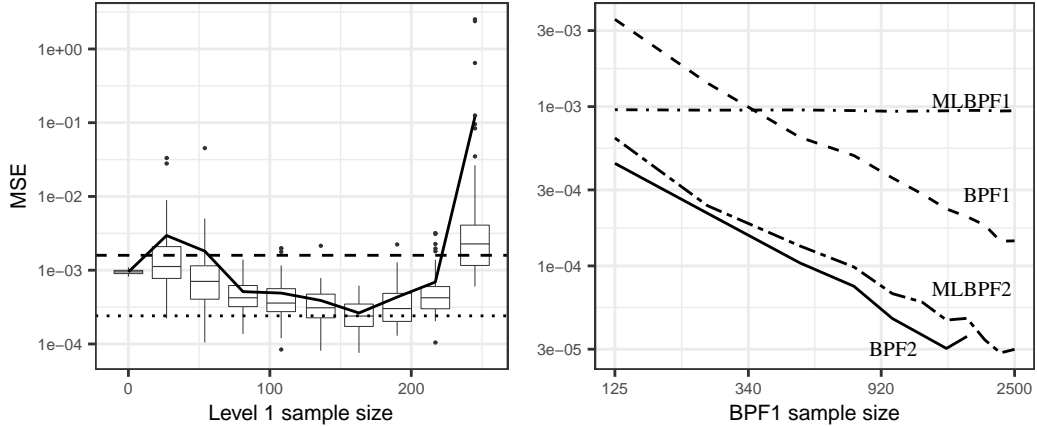}};
\node[align=right,text width = 4cm] at (4.5,1.5) {MLBPF1};
\node[align=right,text width = 4cm] at (4.5,0.25) {BPF1};
\node[align=right,text width = 4cm] at (4.5,-1) {MLBPF2};
\node[align=right,text width = 4cm] at (3.25,-1.8) {BPF2};
\end{tikzpicture}
\caption{Left: Logarithmic MSE boxplots and mean MSE over 50 simulations (solid) for MLBPF with different choices of $(N_0,N_1)$. Dashed line is the mean MSE of the time matched BPF over 50 simulations, dotted line is the mean MSE for error matched BPF. Right: The convergence of BPF and MLBPF as the matching sample size for BPF1 is increased.}
\label{fig:big data}
\end{figure}

The results of the experiment are summarised in Figure \ref{fig:big data}. The left hand side panel shows a logarithmic boxplot of the MSE for MLBPF versus the level 1 sample size $N_1$ based on 50 independent runs with a fixed observation sequence. We have also included the mean RMSE for different sample allocations in Table \ref{tab:time matched sample sizes} from which we see that the best performance of MLBPF is obtained with $(N_0,N_1) = (23664,163)$. In this case, the mean MSE for MLBPF and BPF2 are approximately equal, yet notably smaller than for BPF1. The mean MSE values are: 0.0162 (MLBPF), 0.0399 (BPF1), and 0.0155 (BPF2).

To study the convergence properties of MLBPF, we scaled the sample sizes by  $s\in \{0.5,1,2,3,\ldots 10\}$ and label the MLBPF with $(N_0,N_1) = (68000s,0)$ as MLBPF1, and the MLBPF with $(N_0,N_1) = (23664s,163s)$ as MLBPF2. The sample sizes of BPF1, BPF2 were also scaled similarly, and the resulting mean MSEs are shown in the right hand side panel of Figure \ref{fig:big data} against the corresponding sample size of BPF1 ($250s$). MLBPF2 appears to converge at the same rate with BPF2 (NB: BPF2 results for $s>7$ are omitted due to infeasible computation time). We also observe that MLBPF1 does not converge. This is expected as $N_1=0$ implies that MLBPF1 is essentially a BPF with an incorrect likelihood, and hence will not converge. For this reason in our analysis, we have to assume $c_\ell >0$ for all $0\leq \ell \leq L$. The fact that MLBPF1 does not show any discernible decay either is due to the sample size $N_0$ being relatively large ($=34000$) even for $s=0.5$, so for all $s$ the MSE is approximately constant. We see the same phenomenon in the leftmost boxplot in the left hand side panel of Figure \ref{fig:big data} indicating a very concentrated distribution.

From these results we conclude that with a good choice of $(N_0,N_1)$, MLBPF takes only 14\% of the computation time of BPF to reach the same level of accuracy. To put the scale of the RMSE in the context, it should be noted that the standard deviation (std) of the filtering distribution $\pi_n$ was approximately $0.2429$ throughout the 50 filter iterations. This means that the observed reduction in the RMSE is not negligible nor entirely outweighed by the model uncertainty; by using the MLBPF we are able reduce the error from approximately 16\% of the filter std to only about 7\% of the std --- with the same computational time.

\subsection{Euler-Bernoulli beam}
\label{sec:ebb}

In our second application, we consider the problem of recovering the unknown location of a known moving load on a beam, by observing noisy deflections of the beam at specific points. To make this a filtering problem, the motion of the load on the beam is modelled by a similar signal model as in \eqref{eq:signal model} in Section \ref{sec:big data} with the exception that the initial mean is 1 and signal noise standard deviation $\sigma=0.02$. The beam itself is modelled by the Euler-Bernoulli ordinary differential equation (ODE) model \cite{vinson89}
\begin{align}\label{eq:ebb}
EI\frac{\ud^4}{\ud \ell^4} W_{x} = F_{x}(\ell),
\end{align}
where $E, I \in \R$ are the Young's modulus and the area moment of inertia, that are assumed constant across the beam, and $F_{x}:[0,L] \to [0,\infty)$ is a function representing the load distribution applied to the beam at location $x \in [0,L]$ which is assumed to be known. The solution of this one dimensional ODE is considered across the domain $[0,L]$ denoting the length of the one dimensional beam. Both ends of the beam are assumed to be clamped. 

The deflection of the beam is measured at $m\in\N$ locations $(\ell_{1},\ldots,\ell_{m}) \in [0,L]^{m}$ along the beam subject to Gaussian noise yielding the likelihood proportional to
\begin{align*}
\exp\left (-(y - h^{\theta}(x))^{T}\Sigma^{-1}(y - h^{\theta}(x))/2\right ),
\end{align*}
where $h^{\theta}: [0,L] \to \R^{m}$ is defined as $h^{\theta}(x) = (W_{x}^{\theta}(\ell_{1}),\ldots,W_{x}^{\theta}(\ell_{m}))^{T}$ and  $W_{x}^{\theta}$ is the numerical solution of \eqref{eq:ebb} obtained with solver mesh size $\theta$. For simplicity, we assume the noise to be independent at each measurement location. 

We implemented a two level MLBPF with level specific ODE solver mesh sizes $\theta_0$ and $\theta_1$. Similarly to the example in Section \ref{sec:big data}, by letting $g_{n}^{\theta}$ denote the approximate likelihood obtained with mesh size $\theta$, the level 0 approximation is too poor to be efficiently corrected by the telescoping differences $\Delta g_{n}^{\theta_{1}}$, and therefore the following improved approximation was used.

For level 1, we can write
\begin{align*}
W^{\theta_{1}}_{x}(\ell_{i}) = W^{\theta_{0}}_{x}(\ell_{i}) + \left (W^{\theta_{1}}_{x}(\ell_{i}) - W^{\theta_{0}}_{x}(\ell_{i})\right ), \qquad 1 \leq i \leq m,
\end{align*}
and model the difference $W^{\theta_{1}}_{x}(\ell_{i}) - W^{\theta_{0}}_{x}(\ell_{i})$ by first order linear regression resulting in an approximation 
\begin{align*}
\widetilde{h}^{\theta_{0}}(x) = \begin{pmatrix}W^{\theta_{0}}_{x}(\ell_{1}) + \widehat{\alpha}_1 + \widehat{\beta}_{1}x \\ \vdots \\ W^{\theta_{0}}_{x}(\ell_{m})+ \widehat{\alpha}_{m} + \widehat{\beta}_{m}x\end{pmatrix},
\end{align*}
where the estimates $(\widehat{\alpha}_{1},\widehat{\beta}_{1}),\ldots,(\widehat{\alpha}_{m},\widehat{\beta}_{m})$ can be calculated by simple linear regression using the $N_1$ level 1 particles without notable additional computational cost. 

\begin{remark} Technically, this approximation, as well as the approximation in Section \ref{sec:big data}, is not covered by our analysis due to the dependence of level 0 approximation on level 1 particles. Nevertheless, we conjecture the results in both cases to hold as these approximations are asymptotically independent. Also the experiments appear to confirm our conjecture. 
\end{remark}

The length of the beam was set to  $L=4$ and the observation noise $\Sigma = 0.0002I_{2}$, $I_2$, where $I_2$ is the size 2 identity matrix as the deflection was measured at two locations along the beam $(\ell_1,\ell_2) = (1,1.75)$.

For this application, the exact filter is intractable and therefore a reference BPF with sample size $N=100000$ and mesh size $\theta=4000$ was used as a proxy to the exact filter. To compare the BPF and the MLBPF, we ran a BPF with $N=500$ and $\theta=4000$ (BPF1) and a number of various MLBPF filters with different configurations of level 0 mesh size $\theta_0$ and level specific sample sizes $N_0$ and $N_1$. For each configuration $(\theta_0,N_0,N_1)$, the sample allocation $(N_0,N_1)$ was empirically adjusted to ensure that the computation time of the filer was close to that of BPF1. For each configuration we set $\theta_{1}=4000$. Similarly to Section \ref{sec:big data}, we also run another BPF (BPF2) whose sample size was increased to make the mean MSE over the 50 filter iterations approximately equal to that of the MLBPF with the optimal configuration.

We run the algorithms 10 times for 20 different observation sequences, making the total number of runs 200, for each algorithm. Only 20 observation sequences were used in order to reduce the time spent on running the expensive reference BPF with $N=100000$. This way, it had to be run only 20 times instead of 200.

The results for 200 independent runs are summarised in Table \ref{tab:ebb results} where the BPF filters are regarded as having only level 0 parameters. In this case, the computation time of BPF2 is approximately 6.3 times that of MLBPF with the optimal configuration. Here, the standard deviation of the exact filtering distribution was in the range $[0.004,0.008]$ with mean 0.005, implying that the use of optimally configured MLBPF reduces the error from approximately 14\% of the filter std to about 6\% of the std, with the same computation time.
 
\begin{table}[b]
\caption{Comparison of MLBPF and BPF. Note that BPF1 and BPF2 have only level 0.}
\label{tab:ebb results}

\begin{tabular}{cccccc}
      & $N_0$ & $N_1$ & $\theta_{0}$ & $\theta_1$ & RMSE                  \\\hline
MLBPF & 6133  & 400   & 115          & 4000       &  $3.0\times 10^{-4}$     \\
BPF1  & 500  & -     & 4000         &       -    &  $7.2\times 10^{-4}$     \\
BPF2  & 2000  & -     & 4000         &        -   &  $3.3\times 10^{-4}$     \\
\hline
\end{tabular}

\end{table}

\section{Conclusions}
\label{sec:conclusions}

We have introduced a rigorous definition of the novel multilevel bootstrap particle filter algorithm together with theoretical analysis to establish the strong law of large numbers and central limit theorem. Our numerical experiments suggest great potential in improving the performance from the classical bootstrap particle filter as the computation time in our examples could be reduced to about 15\% of the time required by the classical BPF to reach the same level of error.

While these results are promising, the proposed method is not entirely without concerns. For both examples, the plain vanilla implementation of the MLBPF did not appear to work well and additional application-specific adjustments had to be made. While this is irrelevant to the actual computational efficiency, it does mean that building a generic multipurpose MLBPF algorithm may be challenging as our study appears to suggest that applications have to be considered case by case.

\subsection{Numerical Stability}

We observed that MLBPF was numerically somewhat less stable that the  classical BPF. To some extent, this can been seen in the boxplot of Figure \ref{fig:big data} suggesting that while the mean and the median of the MSE are notably lower than that of the BPF with comparable computation time, the spread of the MSE values across the runs is somewhat alarming and slightly heavy tailed towards large errors. This is evidenced by the mean MSE which is generally larger than the median MSE. We hypothesise this to be due to the situations where there are almost the same number of equally weighted positive particles and negative particles. From our numerical experiments we have also obtained some empirical evidence to support our hypothesis, but a more thorough analysis is left for future work.

It should also be acknowledged that we have not established long term stability i.e.~time uniform convergence for MLBPF nor should we immediately assume this to hold. In the proof of Lemma \ref{lem:as convergence update}, $N_\delta$ depends on $n$ which makes the proof, as such, insufficient for establishing long term stability. Moreover, longer simulations appear to confirm this empirically, as the portion of negative particles tends to grow leading to divisions by approximately zero and hence, unreliable estimates. Nevertheless, we hypothesise long term stability to be achievable by introducing a control on the negative part of the signed measures. Currently such a control mechanism does not exist, but it could potentially be introduced by a modifying the resampling and mutation steps as follows. One can simulate $\xi^{1}_{n+1},\ldots,\xi^{N}_{n+1}$ as an iid sample proportional to the total variation of the signed measure
\begin{align}\label{eq:stability fix}
\dfrac{\sum_{i=1}^{\Ntot(N)}\widetilde{w}_{n}^{i}K(\xi^{i}_{n},\,\cdot\,)}{\sum_{i=1}^{\Ntot(N)}\widetilde{w}_{n}^{i}}.
\end{align}
instead of simulating them as in Algorithm \ref{alg:MLBPF}, whereby $\widehat{\xi}^{1}_{n},\ldots,\widehat{\xi}^{N}_{n}$ and $\xi^{1}_{n+1},\ldots,\xi^{N}_{n+1}$ are drawn proportionally to the \emph{joint} total variation measure where one of the dimensions represents the signed measure components whose linear combination constitutes an approximation for a probability measure. Although technically still a signed measure, the marginal predictive distribution in \eqref{eq:stability fix} is affected only by the net effect of the negative components, instead of the individual negative components of the joint measure. Thus, due to being an approximation of a probability measure, which is unsigned,  \eqref{eq:stability fix} is expected to have a substantially smaller negative part, and therefore, drawing the particles proportionally to the total variation of \eqref{eq:stability fix} is expected to reduce the portion of negative particles.

The downside of sampling from \eqref{eq:stability fix} is its computational complexity. One would have to resort to rejection or importance sampling type methods that require $N$ pointwise evaluations of the density which itself requires $N$ evaluations of the kernel densities making the overall cost $\mathcal{O}(N^2)$. Potentially, the cost could be made more manageable by performing this corrective sampling step only occasionally, analogously to the adaptive resampling strategies~\cite{liu_et_chen98}. A more conclusive assessment of this approach, its feasibility, and ways to reduce the computational cost is an open problem.

\subsection{Complexity theorem and choosing the parameters}

For the classical MLMC, the complexity theorem \cite[Theorem 3.1]{giles08} not only provides conditions under which MLMC provably outperforms classical Monte Carlo, but also gives explicit guidelines on choosing the level-specific sample sizes and the number of approximation levels. We do not have a similar result for MLBPF. Instead, our performance studies are solely based on the numerical experiments, and  the tuning of the algorithm is based on a simple manual search to find the parameter values that yield the best performance as described above in Section \ref{sec:numerical}. 

The reason why we do not have an analogous complexity theorem  immediately available for MLBPF is that in the context of SMC, we consider approximating the \emph{measure} $\widehat{\pi}_n$ rather than a specific integral $\widehat{\pi}_n(\varphi)$ which is the case with the classical MLMC. Fixing the test function $\varphi$ enables one to use the variance or the MSE of the estimate as the optimality criterion, or more generally, as the performance measure that can be used for comparing algorithms. For approximating a measure, variance and MSE are not suitable performance measures and therefore, extending the complexity theorem of \cite{giles08} to MLBPF is not trivial. We hypothesise a more suitable performance measure to be the effective sample size, but the validity of this conjecture will require further analysis.

\begin{appendix}
\section*{Appendix: CLT for triangular martingale arrays}

We use the following Theorem, which is slightly rephrased from the original presentation of \citep[Theorem A.3]{douc_et_moulines08}, to suit our purposes. It is a conditional version of the CLT for 
triangular martingale arrays \citep[Theorem 3.2]{hall_et_heyde80}.

Let $(U_{N,\rho})_{1 \leq \rho \leq \rhomax_{N}}$ be a triangular random variable array such that $\E[U_{N,\rho}\mid \G_{N,\rho-1}] =0$, and let $(\G_{N,\rho})_{0\leq \rho \leq \rhomax_{N}}$ be a triangular array of sub-$\sigma$-algebras of $\F$ of the underlying probability space, such that $\rhomax_{N}$ is $\G_{N,0}$ measurable, and $\G_{N,\rho-1} \subset \G_{N,\rho}$, and for each $N$ and $1 \leq \rho \leq \rhomax_{N}$, $U_{N,\rho}$ is $\G_{N,\rho}$-measurable. Then we have the following result:

\begin{theorem}\label{thm:douc_moulines}
Assume that $\E[U^{2}_{N,\rho}\mid \G_{N,\rho-1}]<\infty$ for all $1 \leq \rho \leq \rhomax_{N}$ and that
\begin{align*}
\begin{array}{rll}
\displaystyle\sum_{\rho=1}^{\rhomax_{N}}\E[U^{2}_{N,\rho}\I[|U_{N,\rho}|\geq \epsilon]\mid \G_{N,\rho-1}] &\inprob{N\to\infty} 0, &\qquad  \text{ for all } \epsilon > 0 \\
\displaystyle\sum_{\rho=1}^{\rhomax_{N}}\E[U^{2}_{N,\rho}\mid \G_{N,\rho-1}] &\inprob{N\to\infty} \sigma^2, &\qquad \text{ for some } \sigma^2>0.
\end{array}
\end{align*}
Then for any $u \in \R$ 
\begin{align}\label{eq:douc_moulines_conclusion}
\E\Bigg[\exp\Bigg(iu\sum_{\rho=1}^{\rhomax_{N}}U_{N,\rho}\Bigg)\,\Bigg|\, \G_{N,0}\Bigg] \inprob{N\to\infty}\exp\bigg(-\frac{u^2}{2}\sigma^2\bigg).
\end{align}
\end{theorem}

\end{appendix}
%
%

\section*{Acknowledgements}
The authors would like to thank Schlumberger Cambridge Research Limited for the financial support for this research. The second author was also supported by EPSRC grant EP/S515279/1.



\bibliographystyle{imsart-number} 
\bibliography{mybib.bib}       


\end{document}